\documentclass[useAMS,usenatbib]{mn2e}

\usepackage{txfonts}
\usepackage[dvips]{graphicx}
\usepackage{epsfig}
\usepackage{natbib}

\usepackage{multirow}
\usepackage{url}
\bibpunct{(}{)}{;}{a}{}{,}
\usepackage{color}
\title[The galaxy halo connection]%
{The galaxy - dark matter halo connection: which galaxy properties are 
correlated with the host halo mass?}

\author[Contreras et al.]
  {
S.~Contreras$^{1,2}$, 
C.~M.~Baugh$^{2}$, 
P.~Norberg$^{2}$, 
N.~Padilla$^{1}$. 
\\
 $^{1}$Instit\'uto Astrof\'{\i}sica, Pontifica
 Universidad Cat\'olica de Chile, Santiago, Chile.\\
 $^{2}$Institute for Computational Cosmology, Department of Physics, 
Durham University, South Road, Durham, DH1 3LE, UK.
}

\def\LaTeX{L\kern-.36em\raise.3ex\hbox{a}\kern-.15em
    T\kern-.1667em\lower.7ex\hbox{E}\kern-.125emX}

\begin{document}

\pagerange{\pageref{firstpage}--\pageref{lastpage}} \pubyear{2014}

\maketitle

\label{firstpage}

\begin{abstract}
We demonstrate how the properties of a galaxy depend on the mass of its   
host dark matter subhalo, using two independent models of galaxy formation. 
For the cases of stellar mass and black hole mass, the median property 
value displays a monotonic dependence on subhalo mass. 
The slope of the relation changes for subhalo masses for which 
heating by active galactic nuclei becomes important. The median property 
values are predicted to be remarkably similar for central and satellite 
galaxies. The two models predict considerable scatter around the median 
property value, though the size of the scatter is model 
dependent. There is only modest evolution with redshift in the median galaxy 
property at a fixed subhalo mass. Properties such as cold gas mass and star 
formation rate, however, are predicted to have a complex dependence 
on subhalo mass. 
In these cases subhalo mass is not a good indicator of the value 
of the galaxy property. We illustrate how the predictions in the 
galaxy property - subhalo mass plane differ from the assumptions 
made in some empirical models of galaxy clustering by reconstructing  
the model output using a basic subhalo abundance matching scheme. 
In its simplest form, abundance matching generally does not reproduce 
the clustering predicted by the models, typically resulting in an 
overprediction of the clustering signal. Using the predictions 
of the galaxy formation model for the correlations between pairs of 
galaxy properties, the basic abundance matching scheme can be extended 
to reproduce the model predictions more faithfully for a wider range 
of galaxy properties. Our results have implications for the analysis 
of galaxy clustering, particularly for low abundance samples. 
\end{abstract} 


\section{Introduction} 
\label{Intro}

How well do different galaxy properties correlate with halo mass? 
Given the value of a galaxy property, such as its stellar 
mass or cold gas mass, how good an indicator is this of the mass 
of the galaxy's dark matter halo? If we know the mass of a dark matter halo in 
a N-body simulation, is there a clear indication of what the properties 
of a galaxy hosted by the halo should be? Here we use two independent 
models of galaxy formation to answer these questions. Our results have 
implications for empirical models which aim to describe 
measurements of galaxy clustering and the construction of galaxy catalogues 
from N-body simulations of structure formation.  

The idea that there should be a connection between the properties of a 
galaxy and the mass of its host dark matter halo lies at the core of 
galaxy formation theory. \citet{WhiteRees:1978} were the first to 
propose that galaxies form when baryons condense inside the 
gravitational potential wells of dark matter halos. The radiative 
cooling of hot gas is just one of the many processes believed to be 
relevant for galaxy evolution (for reviews see \citealt{Baugh:2006} 
and \citealt{Benson:2010}). Even though 35 years that have elapsed 
since the framework for hierarchical galaxy formation was laid down, 
many of the key processes remain poorly understood. Current models use a 
combination of direct simulation and so called ``sub-grid" modelling to 
follow the formation and evolution of galaxies (e.g. \citealt{Cole:2000}; \citealt{Springel:2005}; \citealt{Crain:2009}; \citealt{Schaye:2010}; \citealt{Guo:2011}; \citealt{Vogelsberger:2014}; \citealt{Schaye:2014}). These models now 
give encouraging reproductions of some of the basic characteristics of 
the observed population of galaxies.   

Given the basic tenet laid down by \cite{WhiteRees:1978} it is natural 
that there should be some connection between the mass of a dark 
matter halo and the properties of the galaxy inside it, with the biggest 
galaxies expected to reside in the biggest halos since these halos contain 
the most baryons. This scaling is shaped by feedback processes which 
regulate the rate of star formation. The efficiency of galaxy formation 
varies with halo mass, reaching a peak in halos around the mass of that 
which hosts the Milky Way \citep{EKE:2005,Guo:2010}. In low mass halos, 
heating of the intergalactic medium by photo-ionizing photons and of 
the interstellar medium by supernovae stymie the build up of stellar 
mass \citep{Benson:2002,Somerville:2002}. In high mass halos, 
modellers have appealed to the injection of energy into the 
hot halo by active galactic nuclei (AGN) to reduce the predicted abundance 
of massive galaxies \citep{Benson:2003,Bower:2006,Cattaneo:2006,Croton:2006,Lagos:2008}.  
  
Whilst there is a relation between halo mass and galaxy property for 
some properties, as we will demonstrate, this does not imply that all 
the properties of a galaxy can be deduced once the mass of the host 
halo is specified. Also, the relative importance of the processes which 
take part in galaxy formation varies both with halo mass and redshift. 
This in turn could lead to changes in the manner in which galaxy properties 
scale with halo mass and introduce scatter through a dependence on halo 
formation histories. 

Observed scaling relations between galaxy properties also suggest 
a connection between halo mass and galaxy luminosity 
(see \citealt{Tasitsiomi:2004,Dutton:2010,TG:2011}). 
\cite{T&F:1977} found a tight correlation between galaxy luminosity, $L$, 
and the circular velocity of the disk, $V_{\rm c}$, for spiral galaxies.  
In the optical, the scaling is $L \propto V_{\rm c}^{3}$ 
\citep{Mocz:2012}. In the near-infrared this becomes 
$L \propto V_{\rm c}^{4}$ \citep{Verheijen:1997,Tully:1998}. 
A similar scaling exists for elliptical galaxies, albeit with a larger 
scatter \citep{F&J:1976}.

It is tempting to use these observed galaxy scaling relations to assign 
a luminosity to a dark matter structure with a given circular velocity. 
However, there are number of problems with such an approach. 
First, the precise scaling relation depends on the galaxy selection, with 
different scalings found for spirals and ellipticals. 
Second, the observed relations only cover a limited dynamic range 
in circular velocity and luminosity, and so cannot be applied to low 
mass halos. Finally, the application 
of the scaling relation assumes that the circular velocity measured for 
the galaxy can easily be related to the circular velocity which 
characterizes the dark matter halo, whereas in reality these are measured 
at very different radii. Models suggest that shifts of 20-30\% are 
common between the circular velocity at the half-light radius of the 
galaxy and that obtained at the virial radius of the halo 
(e.g. \citealt{Cole:2000}). This difference in velocity would make a 
big difference to the assigned galaxy luminosity, given the steep 
dependence of the observed scaling relations on circular velocity.  

A more promising approach to connect galaxies with their host dark matter 
halos is the sub-halo abundance matching (SHAM) technique introduced 
by \cite{ValeOstriker:2004}, who proposed a monotonic relation between 
galaxy luminosity and halo mass with zero scatter (e.g. \citealt{Kravtsov:2004,
Vale:2006,Conroy:2006}; for a review of galaxy clustering models 
see \citealt{Baugh:2013}). A galaxy catalogue with spatial information 
can be constructed using SHAM by taking a sample of galaxy luminosities, 
generated, for example, using an observed galaxy luminosity function, 
sorting in luminosity and then matching up this list of galaxies 
with a sorted list of subhalo masses obtained from an N-body simulation.    
The SHAM technique has been used extensively to model galaxy clustering 
(eg. \citealt{Conroy:2006,Shankar:2006,Baldry:2008,Moster:2010,Guo:2010,Behroozi:2010,Wake:2011,Hearin:2013,Nuza:2013,Reddick:2013,SimhaCole:2013}).

The modern implementation of SHAM has one important difference 
from the original proposal of \cite{ValeOstriker:2004}. This regards 
the treatment of satellite galaxies. These galaxies reside in dark 
matter structures called subhalos which may have experienced significant mass 
loss, depending on their orbit within their more massive dark matter halo. 
Using the instantaneous subhalo mass measured from a N-body simulation 
would therefore lead to an error in the assigned luminosity. 
To circumvent this, the mass of the subhalo at the point of infall to 
the larger structure is commonly used \citep{Conroy:2006,Vale:2006}. We 
note that recent N-body simulations have shown that the maximum halo mass 
is attained prior to infall, with some mass loss already occuring before 
the halo crosses the virial radius of the more massive halo 
\citep{Behroozi:2014}. Furthermore, some satellite galaxies should be assigned 
to subhalos which can no longer be identified in a given simulation output 
due to the finite resolution. The issue of identifying a suitable dark 
matter structure to assign a galaxy to can be avoided if multiple 
outputs are available and the formation history of subhalos can be 
extracted (\citealt{Conroy:2006,Conroy:2009}; see also \citealt{Klypin:2013} 
and \citealt{Guo:2014} for requirements on the resolution of subhalos).  

The original SHAM proposal relies on two key assumptions: (i) 
there is zero scatter between the galaxy property and halo mass, 
(ii) the impact of environmental effects on galaxy properties can be ignored. 
We will show that the first assumption is not supported by 
current galaxy formation models. The second assumption is also 
 not held in most galaxy formation models, which explicitly treat 
gas cooling onto satellites and centrals differently 
(but see \citealt{Font:2008} and \citealt{Guo:2011} for alternative models). 
Hydrodynamic simulations show that this distinction may be blurred, with gas 
cooling continuing onto satellite galaxies \citep{McCarthy:2008,Simha:2009}. 
Observationally, the environment is found to shape the properties of galaxies 
\citep{Balogh:2004,Peng:2010}. 

Even though the basic SHAM model is still discussed extensively in 
the literature (e.g. to give just two recent examples \cite{Finkelstein:2015,Yamamoto:2015}), we note that various extensions to the model have been proposed 
which try to account for scatter in the value of a galaxy property associated 
with a given subhalo mass \citep{Tasitsiomi:2004,Behroozi:2010,Moster:2010,
Neistein:2011,Reddick:2013} and which assign galaxy properties that do not 
have a simple dependence on halo mass \citep{RP:2011,Hearin:2013,Gerke:2013,Masaki:2013,Kravtsov:2014,Hearin:2014,RP:2014}. 
 
Here we examine the nature of the galaxy - halo connection 
using semi-analytic galaxy formation models (SAMs). These models 
represent a physically motivated, ab-initio calculation which tracks 
the fate of the baryonic content of the Universe. SAMs naturally 
predict the number and properties of galaxies in dark matter halos 
as a function of halo mass. \cite{Simha:2012} carried out a similar 
analysis using smoothed particle hydrodynamics simulations. 
These simulations were run using small computational volumes and 
so did not include AGN feedback, which meant that the high mass end 
of the stellar mass - halo mass relation could not be studied. 
One advantage of using SAMs is that they can be run using the dark 
matter halo merger trees from N-body simulations covering different 
volumes and mass resolutions, allowing a very wide dynamic range of 
mass to be probed at a low computational cost. To establish the robustness 
of the model predictions, we use two SAMs from independent groups: 
one which uses {\tt GALFORM} (\citealt{Lagos:2012}) and the other 
which uses the {\tt L-GALAXIES} code (\citealt{Guo:2011}). These models are 
representative of the current state-of-the-art of semi-analytical modelling.

The main aim of our paper is to establish which galaxy properties 
show a simple dependence on subhalo mass and how much scatter there is 
in the value of a galaxy property for a given halo mass. We consider 
the intrinsic galaxy properties of stellar mass, cold gas mass, 
star formation rate and black hole mass. We also study luminosities at 
different wavelengths, ranging from the ultra-violet, which is sensitive 
to the recent star formation history of a galaxy, to the near-infrared, 
which correlates more closely with its stellar mass. 
To illustrate the features of the model predictions, we compare 
the output of the galaxy formation model to some simple 
empirical models of galaxy clustering.
We do this by applying the original, baisc SHAM model
to reconstruct the SAM catalogues, comparing the clustering measured 
from the reconstructed catalogue with the prediction from the original 
catalogue. 
Taking advantage of the galaxy formation output, which tells us how 
different galaxy properties are correlated, we also consider a 
simple ``two-step" SHAM approach for properties which do not 
meet the SHAM hypothesis themselves (see e.g. \citealt{RP:2011}). This 
also allows us to include at some level the scatter in the galaxy 
property - subhalo mass relation (see \citealt{Trujillo-Gomez:2011, Hearin:2013, Masaki:2013} for more detailed discussion of models with similar aims). 
A key advantage of our study is that we extract the subhalo mass at infall 
into a more massive halo using the halo merger trees which are used 
in the semi-analytical model. This means that the problem 
of ``missing subhalos" that afflicts SHAM when applied to a single 
N-body output is not an issue.   

Our earlier paper comparing the clustering predictions made 
by different SAMs shows that the models are sufficiently 
robust for the exercise carried out here \citep{C13}. 
For galaxy samples selected by stellar mass, the {\tt L-GALAXIES} and 
{\tt GALFORM} models make remarkably similar clustering predictions on large 
scales. There are differences in the clustering predicted on 
small scales, but \cite{C13} show how these can be understood in 
terms of choices made in the implementation of galaxy mergers 
(see Section~\ref{subsec:mergers} for further discussion). 

The layout of the paper is as follows. In Section~2 we first introduce 
the two semi-analytical models of galaxy formation used (\S~2.1) and 
the N-body simulations they are implemented in (\S~2.2). The definition 
and identification of subhalos is discussed in \S~2.3; subhalos also 
play a role in galaxy mergers, as set out in \S~2.4. The resolution 
ranges of the predictions, in terms of subhalo mass and galaxy properties 
is covered in \S~3. The main results are presented in \S~4, where 
we present model predictions for how galaxy properties depend on subhalo 
mass (\S~4.1), show which halos contribute to galaxy samples when different 
selections are applied (\S~4.2) and illustrate what happens when SHAM is 
used to reconstruct the theoretical models (\S~4.3). Our results are 
summarized and presented along with our conclusions in \S~5.

\section{The galaxy formation models}
Here we give a brief overview of the galaxy formation models 
used in our study along with the specifications of the N-body 
simulations they are grafted onto. In Section~\ref{subsec:SAM} we briefly 
introduce the two SAMs and list the physical process they 
attempt to model. In Section~\ref{subsec:Nbody}
we describe the dark matter simulations in which both SAMs 
are implemented. The definitions of subhalo mass used in the two models 
is discussed in Section~\ref{subsec:DMS}. Finally, in Section~2.4 
we list the steps necessary to be able to compare models which employ 
different definitions of subhalo mass.

\subsection{Semi-analytic models}
\label{subsec:SAM}
The SAMs used in our comparison are those of \cite{Lagos:2012} (hereafter L12) 
and \cite{Guo:2011} (henceforth G11) \footnote{The G11 outputs are publicly 
available from the Millennium Archive in Garching \url{http://gavo.mpa-garching.mpg.de/Millennium/}}.

The objective of SAMs is to model the main physical processes 
involved in galaxy formation and evolution in a cosmological context: 
(i) the collapse and merging of dark matter halos; 
(ii) the shock heating and radiative cooling of gas inside dark matter 
halos, leading to the formation of galaxy discs; (iii) quiescent star 
formation in galaxy discs; (iv) feedback from supernovae (SNe), from accretion 
of mass onto supermassive black holes and from photoionization heating of 
the intergalactic medium (IGM); (v) chemical enrichment of the stars and gas; 
(vi) dynamically unstable discs; (vii) galaxy mergers driven by dynamical 
friction within dark matter halos, leading to the formation of stellar 
spheroids, which may also trigger bursts of star formation. 
The two models have different implementations of each of these processes. 
By comparing models from different groups we can get a feel for 
which predictions are robust and which depend on the particular 
implementation of the physics. 

The G11 model is based on various models from the Munich group 
\citep{DeLucia:2004,Croton:2006,DeLucia:2007}. The L12 model is a 
development of the model of \cite{Bower:2006} which 
includes AGN heating of the cooling gas in massive halos. The L12 model 
has an improved treatment of star formation, breaking the interstellar 
medium into molecular and atomic hydrogen components \citep{Lagos:2011b}. 
One important difference between G11 and L12 is the implementation 
of cooling in satellite galaxies. In L12, a galaxy is assumed to lose 
its hot gas halo completely once it becomes a satellite; in G11, this 
process is more gradual and depends on the orbit of the satellite. 
Another important difference is the treatment of galaxy mergers. 
This will be discussed in Section~\ref{subsec:mergers} after we have 
introduced the N-body simulations used and the dark matter halo catalogues 
derived from them. 

\begin{table}
\begin{center}
    \begin{tabular}{| l | l | l | c |}
    \hline
    Simulation & $ N_{\rm P}$ & $ m_{\rm P}/h^{-1} {\rm M_{\odot}}$ & $ L/ h^{-1}{\rm Mpc}$\\ \hline
    MS-I & $2160^3$ & $8.61 \times 10^8$& 500 \\ \hline
    MS-II & $2160^3$ & $6.88 \times 10^6$& 100 \\ \hline
    \hline
    \end{tabular}
\caption{The numerical parameters of the N-body simulations used. MS-I is the N-body simulation of \citet{Springel:2005} and MS-II is the simulation described by \citet{BK:2009}.} 
\end{center}
\end{table}

\subsection{N-body simulations}
\label{subsec:Nbody}
The SAMs used in this paper are both implemented in two N-body simulations, 
the Millennium I simulation (\citealt{Springel:2005}, hereafter MS-I) 
and the Millennium II simulation (\citealt{BK:2009}, MS-II from now on). 
The  properties of the simulations are listed in Table~1. These two 
simulations have the same cosmology\footnote{The values of the 
cosmological parameters used in the MS-I \& -II are: 
$\Omega_{\rm b}$ =0.045, $\Omega_{\rm M}$ = 0.25, $\Omega_{\Lambda}$ = 0.75, 
h = $H_0/100$ = 0.73, $n_{\rm s}$ = 1, $\sigma_8$ = 0.9.} and the 
same number of particles, but employ different volumes and hence have 
different mass resolutions. There are 63 and 67 simulation outputs between 
$z=127$ and $z=0$ for MS-I and MS-II respectively. Halo finding algorithms were run on these outputs and 
used to build halo merger trees, as outlined in the next section. 
These trees are the starting point for the SAMs. 
By implementing the SAMs in different volume simulations, we can 
study the model predictions over a much wider range of halo mass 
than would be possible with a single simulation. 

\subsection{Dark matter subhalos}
\label{subsec:DMS}
Once a halo becomes part of a more massive structure it is called a 
subhalo. The subhalo can retain its identity for some time after becoming 
gravitationally bound to the larger halo. Tidal forces lead to the 
removal of mass from the subhalo. The extent of this mass ``stripping" 
depends upon the orbit followed by the satellite, with the tidal forces 
being stronger closer to the centre of the main halo. Dynamical friction 
will also cause the orbit of the subhalo to decay, moving the subhalo 
closer to the centre of the halo. 

Friends-of-Friends ({\tt FoF}) groups \citep{Davis:1985} are identified 
in each simulation output and retained down to 20 particles. 
{\tt SUBFIND} is run on these groups to identify subhalos within 
the {\tt FoF} groups \citep{Springel:2001}. The construction of the 
dark matter halo merger histories using this information differs 
from this point onwards between the two groups (for further details of 
the merger tree construction, see \citealt{Guo:2011} and \citealt{Jiang:2014}).

Eventually, if the mass-stripping is severe, {\tt SUBFIND} will no longer 
be able to locate the subhalo. This poses a problem when attempting 
to apply SHAM to a single output from a N-body simulation. If many 
outputs are available, however, it is possible to build halo merger trees 
and to track the subhalo until {\tt SUBFIND} is unable to locate it; 
thereafter the location of the galaxy associated with the subhalo is typically 
assigned to the potential minimum of its subhalo \citep{Jiang:2014}. 

As a result of the mass stripping experienced by subhalos, neither the 
instantaneous mass nor the maximum effective circular velocity of the 
halo rotation curves are useful indicators of the subhalo mass prior 
to infall \citep{Ghigna:1998,Kravtsov:2004}. \citet{Conroy:2006} proposed 
that the mass of the subhalo at infall should be used instead as a more 
reliable measure of the subhalo mass, using the effective maximum circular 
velocity as a proxy (see also \citealt{Vale:2006}). 

Here, we use the mass of the subhalo at the point of infall into a 
larger structure as obtained from the halo merger history if the host 
galaxy is a satellite, or the current halo mass if the galaxy is a central. 
Throughout the paper we will refer to the subhalo mass at infall as the subhalo 
mass unless explicitly stated otherwise.

The subhalo mass is obtained from the halo merger history, which  
is constructed using independent algorithms by the Durham and Munich 
groups. G11 construct dark matter halo merger trees by first running a 
FoF percolation algorithm on each simulation output or snapshot. 
{\tt SUBFIND} is then run on the {\tt FoF} halos to identify the bound 
particles and substructures within the halo. The merger tree is constructed by linking a subhalo in 
one output to a unique descendant subhalo in the subsequent snapshot. 
The halo merger tree used in the Munich SAM is therefore a subhalo 
merger tree. The L12 SAM uses the {\tt DHalos} merger tree construction 
(\citealt{Jiang:2014}; see also \citealt{Merson:2013} and 
\citealt{Gonzalez-Perez:2014}). The initial steps are the same as in 
the Munich case, running {\tt FoF} and {\tt SUBFIND} on the simulation 
outputs. Additional considerations are applied in the construction 
of the {\tt DHalo} merger trees. These include the requirement of the 
Durham SAM that halo mass increases monotonically with the age of 
the universe and the analysis of the halo at future snapshots to avoid 
the premature linking of halos which pass through another halo. 
The relation between subhalo masses in the L-GALAXIES and GALFORM 
cases is composed of an offset in mass and a scatter 
\citep{Jiang:2014}. In~\S3.1 we will come up with a simple scheme 
to relate halo masses in the two SAMs.

\subsection{Galaxy mergers}
\label{subsec:mergers}

SAMs generally distinguish between two classes of satellite galaxies, type-I satellites 
which are associated with resolved DM subhalos and type-II satellites, 
also called ``orphans", for which the host subhalo can no longer be 
identified by {\tt SUBFIND}. In {\tt L-GALAXIES}, this information is 
used to decide which galaxies are candidates to merge with the central 
galaxy in the halo. Satellite galaxies which are associated 
with a resolved subhalo, i.e. type-I galaxies, are not allowed to 
merge with the central galaxy in their host dark matter halo. 
Once sufficient stripping of the dark matter has occured, such that the 
host subhalo can no longer be resolved and the type-I galaxy has become a type-II, a dynamical friction timescale is calculated for the galaxy to merge 
with the central. In the {\tt GALFORM} model studied here, the presence 
of the subhalo is ignored for this purpose and all satellite galaxies 
are considered as candidates to merge with the central galaxy and a 
dynamical friction timescale is calculated for each satellite. This choice 
leads to a difference in the small-scale clustering predicted by the 
{\tt L-GALAXIES} and {\tt GALFORM} models, even in the case when the 
models contain the same number of satellites, as the radial distribution 
of satellites is different \citep{C13}. We note that in the current version 
of {\tt GALFORM} it is possible to select a galaxy merger scheme that 
operates in the same fashion as the one used 
in {\tt L-GALAXIES} \citep{Campbell:2014}. 

\begin{figure}
\includegraphics[bb= 22 151 346 704, width=0.48\textwidth]{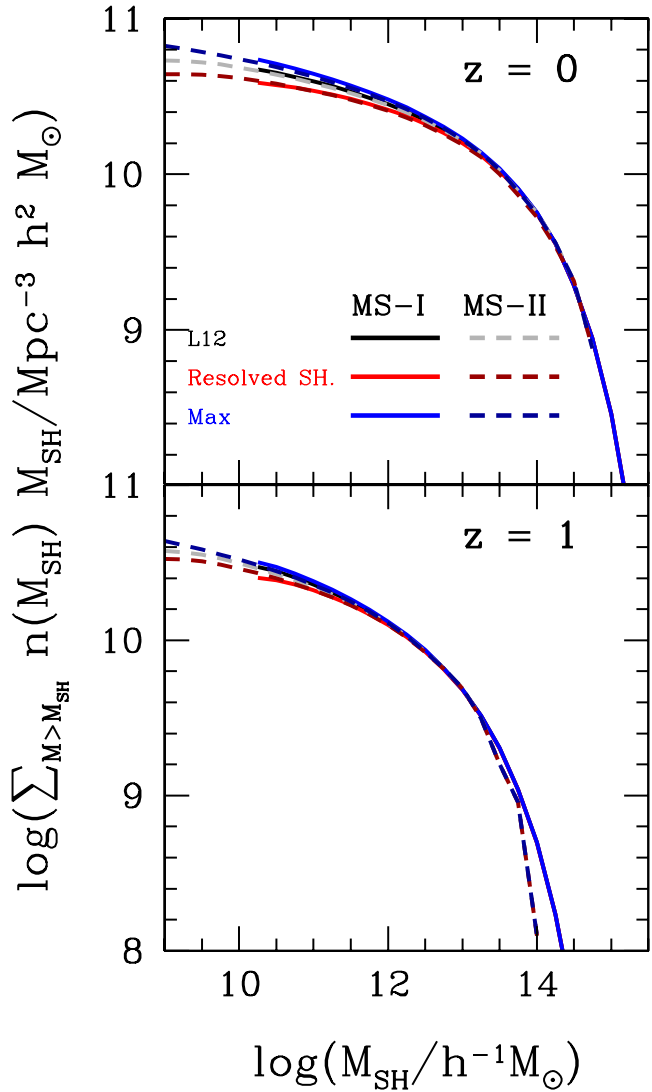}
\caption{
The cumulative mass contained in subhalos in the L12 model at $z=0$ 
(top) and $z=1$ (bottom).  
The solid lines show predictions from the MS-I and the dashed lines 
from MS-II. The black curves show the predictions using both 
resolved and unresolved subhalos (as obtained from the halo merger tree; 
see text). The red curves 
show the results for resolved subhalos only. The blue curves 
show the predictions for a model in which the number of subhalos 
is maximized by, effectively, allowing all halos to cool gas 
efficiently by removing stellar and photoionization feedback and  
switching off galaxy mergers. 
}
\label{fig:SHMF1} 
\end{figure}

\section{Resolution limits of the SAMs} 

In this section we explain how we determine the range of subhalo 
masses and galaxy properties over which we consider the results 
obtained from the MS-I and MS-II simulations. Section~\S3.1 discusses 
the subhalo mass function and \S3.2 presents the limits for the 
different galaxy properties.

\begin{figure}
\includegraphics[bb= 22 151 346 704, width=0.48\textwidth]{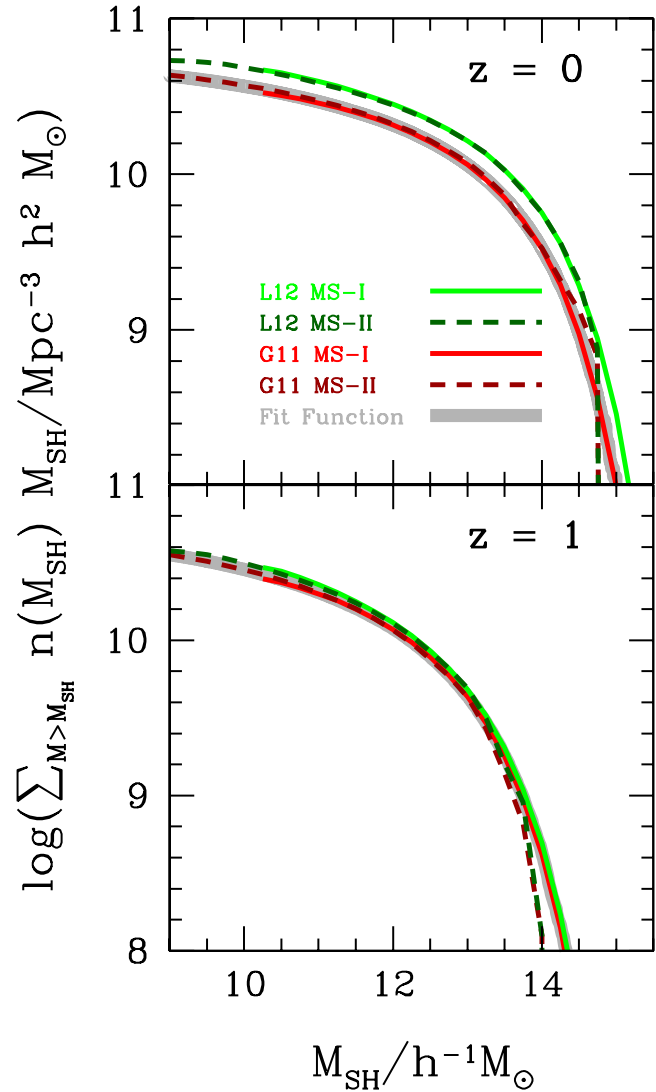}
\caption{The distribution of subhalo masses in the L12 and G11 models, 
using the MS-I (solid lines) and MS-II (dashed lines), as labelled in 
the top panel, at $z=0$ (top) and $z=1$ (bottom). The thick grey line 
shows a fitting function which matches the subhalo mass function 
in the G11 model from MS-I for subhalos more massive than 
$10^{11} h^{-1} {\rm M_{\odot}}$ and from the MS-II for less massive halos. 
From now on, the subhalo masses quoted for both SAMs will be rescaled 
with reference to this curve, such that the predicted subhalo mass functions 
coincide with the fitting function.
}
\label{fig:SHMF2} 
\end{figure}

\subsection{The subhalo mass function}

The cumulative distribution of subhalos masses in the L12 model is shown 
in Fig.~\ref{fig:SHMF1}, in which we plot the total mass contained
in subhalos with masses greater than $M_{\rm sh}$, $\int_{M_{\rm sh}}^{\infty}  
n_{sh}(M) M {\rm d}M$, at $z= 0$ (top) and $z = 1$ (bottom). 
Due to the way in which we construct the subhalo mass function by using 
galaxies to point to their host subhalo, the subhalo mass function is nominally 
dependent on the galaxy formation model used. In the case of central galaxies, the 
mass of the host halo is used. For satellite galaxies we always use the mass of 
the host halo at the time of infall into a more massive structure. This 
information is obtained from the galaxy merger history predicted by the SAM. 

The number of galaxies output by the SAM can change if, for example, 
the heating of the intergalactic medium by photoionization varies or the 
rate at which galaxies merge is altered. To explore the dependence 
of the subhalo mass function on galaxy formation physics, 
we have run an extreme variant of the L12 model in which we have deliberately 
set out to maximize the number of galaxies and, consequently, 
the number of subhalos picked up from the dark matter halo merger trees. 
This model has a cooling time set to zero in all halos, has no supernova 
feedback and has a galaxy merger timescale that is set to infinity. This means 
that galaxies will form in all subhalos and will not be removed by mergers. 
The mass in subhalos in this variant model is shown by the 
blue curves in Fig.~\ref{fig:SHMF1}. The agreement with the 
predictions using the standard L12 model is impressive; the 
subhalo mass functions are indistinguishable at $z=0$ above 
a subhalo mass of $10^{12} h^{-1} {\rm M_{\odot}}$, and only differ by 
up to around 50\% at lower masses.  

The results from the MS-I and MS-II simulations overlap reasonably well, 
with the MS-II predictions extending to lower subhalo masses and displaying 
more noise at the high-mass end due to the smaller simulation volume. 
The black line in Fig.~\ref{fig:SHMF1} shows the mass in subhalos associated 
with all galaxies (i.e. for type-II galaxies without a resolved subhalo, 
we use the subhalo mass at infall), whereas the red curve shows how this 
mass is reduced when only galaxies attached to resolved subhalos are 
considered. 

We now compare in Fig.~\ref{fig:SHMF2} the subhalo mass functions 
obtained from the L12 and G11 SAMs. One difference between the subhalo masses 
reported by the two groups is that the DHalo mass used in 
{\tt GALFORM} corresponds to an integer number of particles whereas 
a virial mass is calculated in {\tt L-GALAXIES}. Hence, the G11 subhalo 
masses can extend down to lower masses than in the L12 case. The G11 
subhalo masses can also decrease over time, unlike the {\tt DHalo} masses, 
which, by construction, increase monotonically. 

To enable us to plot galaxy properties against subhalo mass 
and to compare the two models using the MS-I and MS-II simulations, 
we need to take into account the offset in the predicted subhalo mass 
functions, as plotted in Fig.~\ref{fig:SHMF2}, which is due to the 
differences mentioned above in the definition of halo mass. We do this 
by defining a smooth function which describes the form of the subhalo 
mass function.\footnote{We note that \cite{Jiang:2014} show that the 
halo masses used in {\tt GALFORM} and {\tt L-GALAXIES} 
are related by an offset with 
a scatter.} We force this function to fit the subhalo mass function 
of the G11 model using the MS-I for masses above 
$10^{11} h^{-1} {\rm M_{\odot}}$. 
For halos less massive than this value, we use the G11 subhalo mass 
function from MS-II. The L12 subhalo masses are effectively rescaled, so 
that for a given subhalo abundance, the subhalo mass is that derived 
from the smooth fitting function at the same space density of objects.  

The original subhalo mass functions of L12 and G11 run 
with MS-I and MS-II are shown in Fig.~\ref{fig:SHMF2} for 
$z = 0$ and $z = 1$. The differences in the subhalo mass functions 
in the two models are clearly visible and depend on redshift. 
The smooth fitting function derived from the combination of G11 run 
with the MS-I and MS-II is shown as a thick grey line. From now on, 
all the SAM predictions will use this subhalo mass definition.

\begin{figure}
\includegraphics[bb= 22 151 576 666,width=0.48\textwidth]{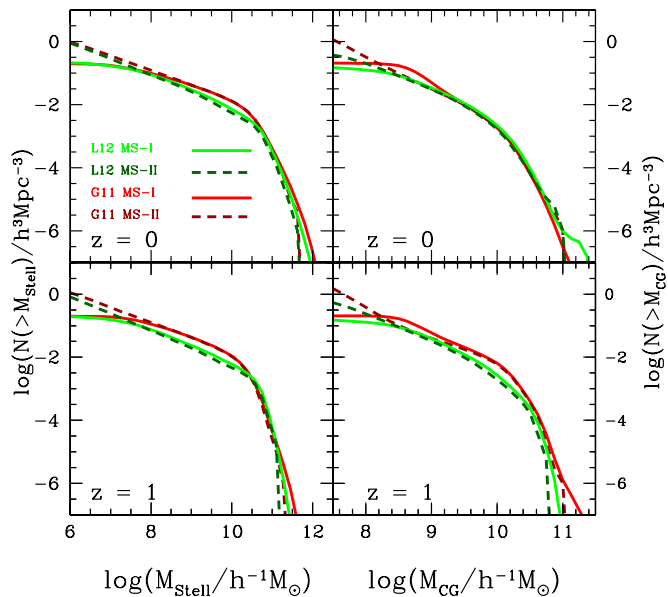}
\caption{Cumulative stellar mass (left panels) and cold gas mass 
(right panels) functions for $z = 0$ (top) and $z = 1$ (bottom), 
for the L12 and G11 models, as labelled, obtained 
from the MS-I (solid lines) and MS-II (dashed lines).}
\label{fig:StellCG} 
\end{figure}

\subsection{Galaxy properties}

The distributions of galaxy properties predicted by the models are more 
complex than those of subhalos. One issue is that for some properties, 
such as black hole mass or star formation rate, some galaxies are predicted 
to have zero values. The fraction of galaxies with zero values for a 
particular property can vary strongly between models. Hence, we do not 
attempt to replicate the approach taken for subhalo masses in the previous 
section. Instead we determine the range of property values to use from 
the MS-I and MS-II runs for each model separately. 

The distribution of cold gas masses and stellar masses predicted by the SAMs 
is plotted in Fig.~\ref{fig:StellCG}. Whilst there is, reassuringly, reasonable 
agreement between the predictions of a given model for 
the MS-I and MS-II runs for intermediate property values, there are clear 
differences between the L12 and G11 models. 
This is to be expected given the differences in the way in which the model 
parameters are calibrated and in choices such as the stellar initial mass 
function and the stellar population synthesis model used to convert the 
predicted star formation histories into luminosities. 

We use the galaxy properties predicted using the MS-I run for galaxies with 
larger values of properties such as stellar mass or cold gas mass. Moving 
in the direction of smaller property values, once the cumulative distribution 
obtained from MS-I differs from that recovered using MS-II by more than a 
given amount, we switch to using the higher mass resolution MS-I results. 
Where practicable, we set the tolerance between the mass functions to 
be 5\% before switching over to the MS-II predictions. Combined with the 
overall differences between the model predictions, this means that the 
transition between the MS-I and MS-II predictions is made at different 
property values for each model. To compare models, we set a number density 
to define galaxy samples, and select property values in each model to attain 
this number density.

\begin{figure}
\includegraphics[width=0.48\textwidth]{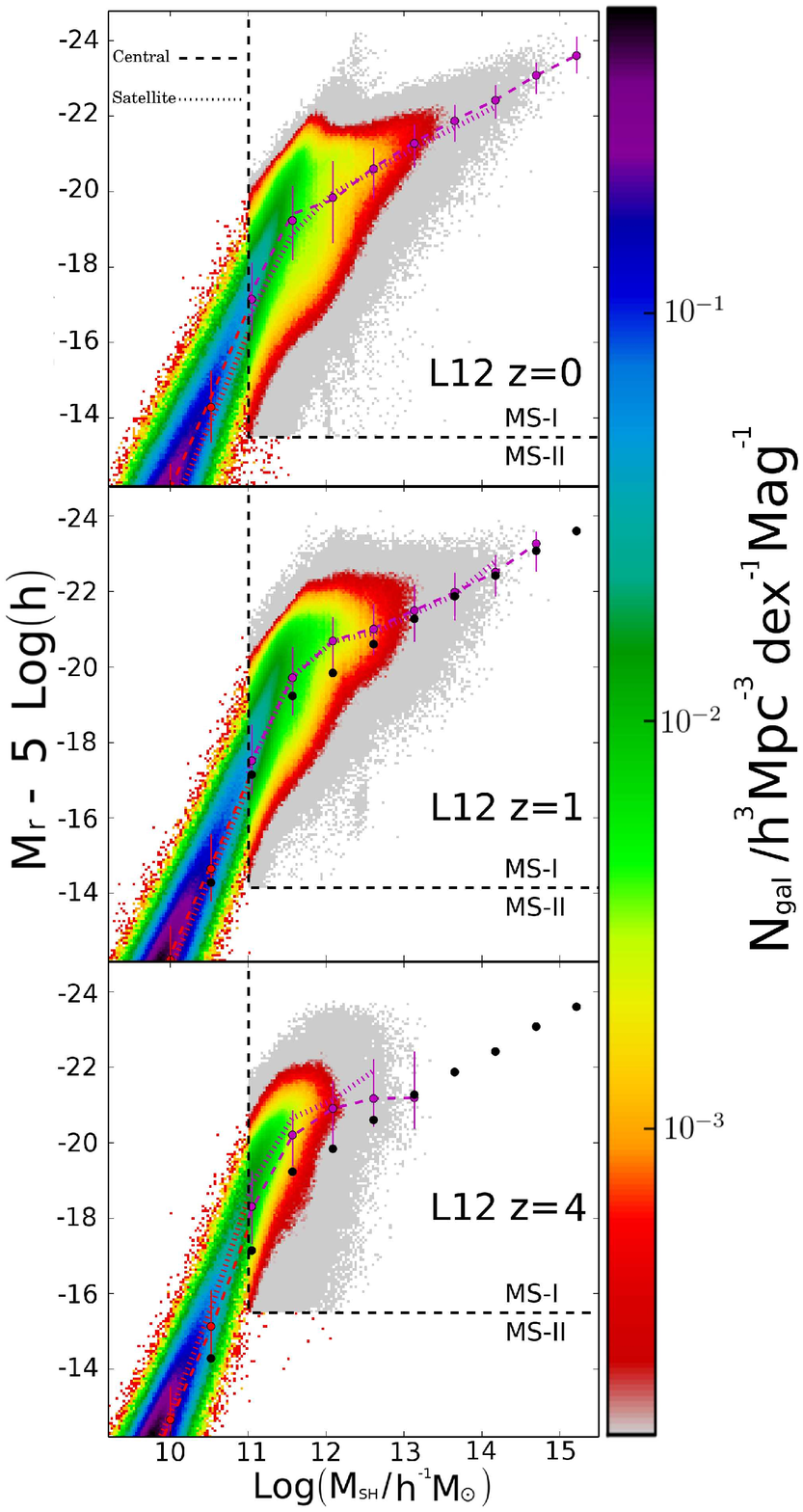}
\caption{
The distribution of the rest-frame $r$-band magnitude as a function of subhalo 
mass predicted in the L12 model, at $z = 0$ (top), $z = 1$ (middle) and 
$z = 4$ (bottom). The colour shading represents the space density of galaxies 
as indicated by the colour bar on the right. The symbols with error bars 
show the median $r$-band magnitude and the 20-80$^{\rm th}$ percentile range 
for all galaxies. The black dots in the $z = 1$ and $z = 4$ panels show the 
median of the $r$-band magnitude at $z = 0$, which is reproduced in these 
panels for reference. Different line styles show the 
median relation for centrals (dashed lines) and satellites (dotted lines) 
separately. The dashed line box separates the MS-I predictions (top right 
region) from those obtained from the MS-II, where the cumulative luminosity
functions from the MS-I and MS-II differ by more than 5\%. 
}
\label{fig:2D_Mr} 
\end{figure}

\begin{figure*}
\includegraphics[width=1.0\textwidth]{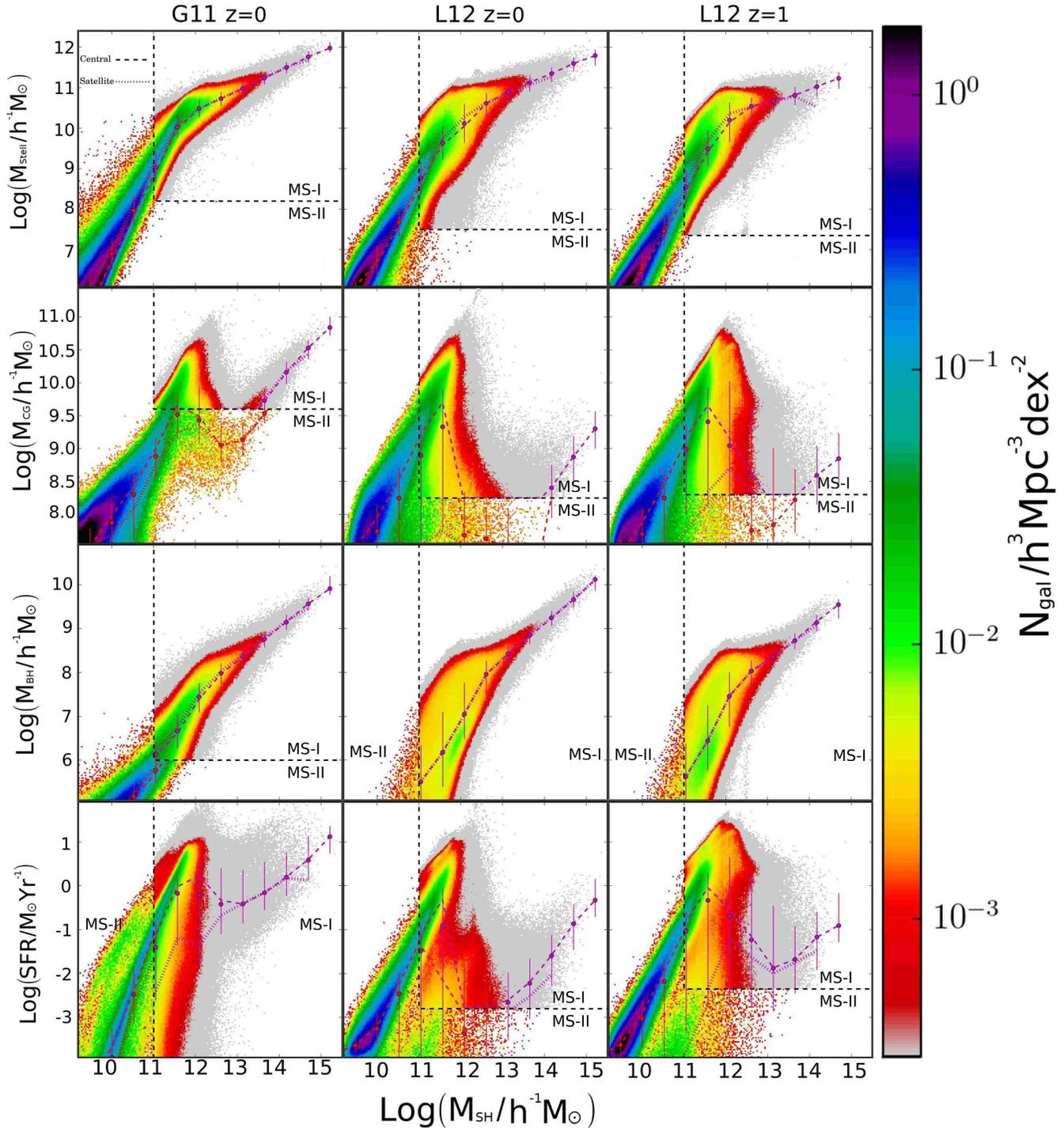}
\caption{The predicted distributions of physical galaxy properties with 
subhalo mass (stellar mass, top row; cold gas mass, second row; 
black hole mass, third row; star formation rate, bottom row). 
The first column shows the G11 model at $z=0$. The second and third 
columns show the L12 model at $z=0$ and $z=1$ respectively. The colour 
shading shows the number density of galaxies as indicated by the key on 
the right. The black dashed lines show the transition from the MS-I to MS-II 
predictions and are in different places for the two models. The points with 
error bars show the median property values and the 20-80$^{\rm th}$ percentile 
range. The median property values are also shown for central (dashed lines) 
and satellites (dotted lines).}
\label{fig:2D} 
\end{figure*}

\section{Results}
\begin{figure}
\includegraphics[bb=30 164 577 700,width=0.47\textwidth]{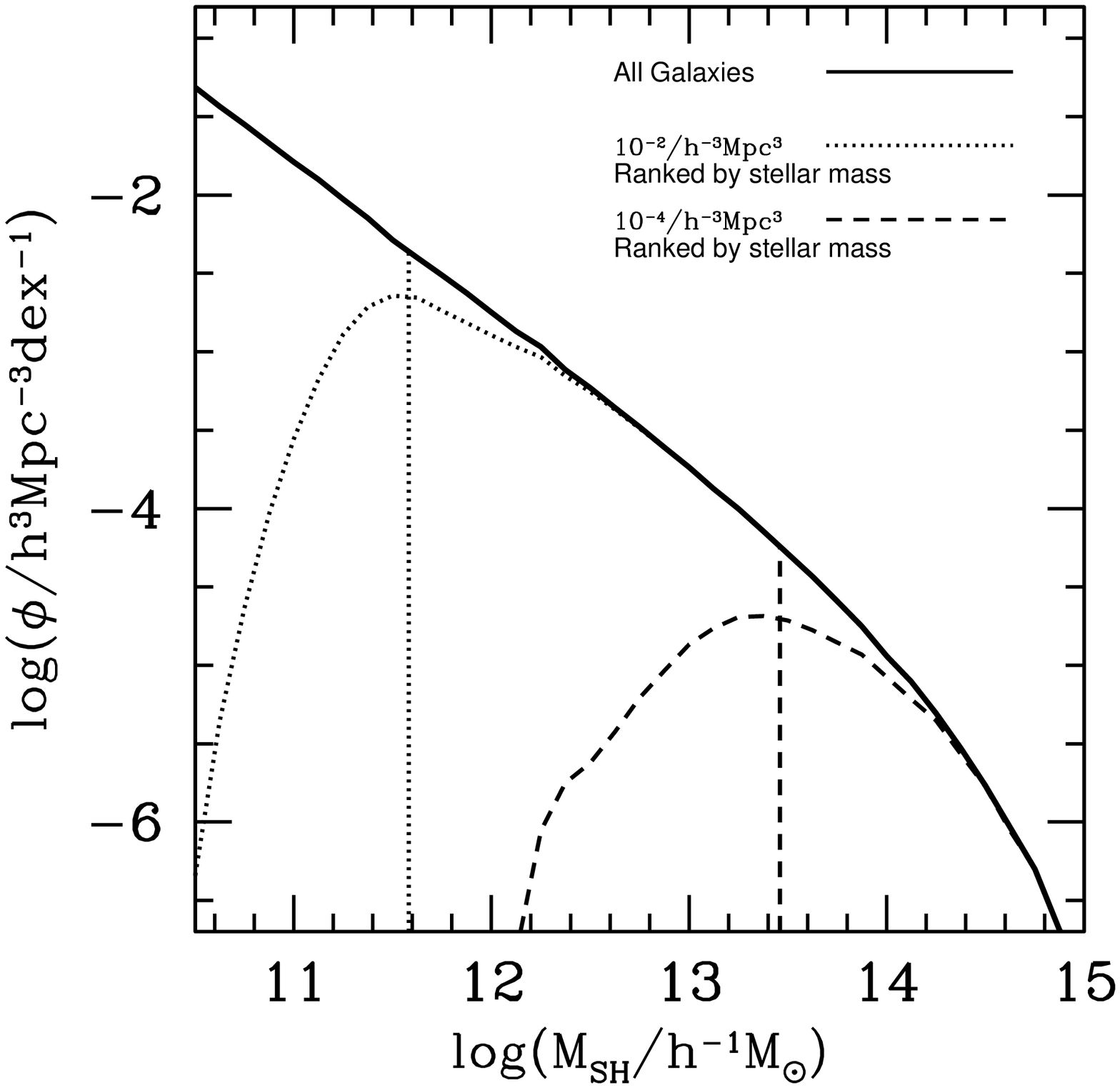}
\includegraphics[bb=30 164 577 700,width=0.47\textwidth]{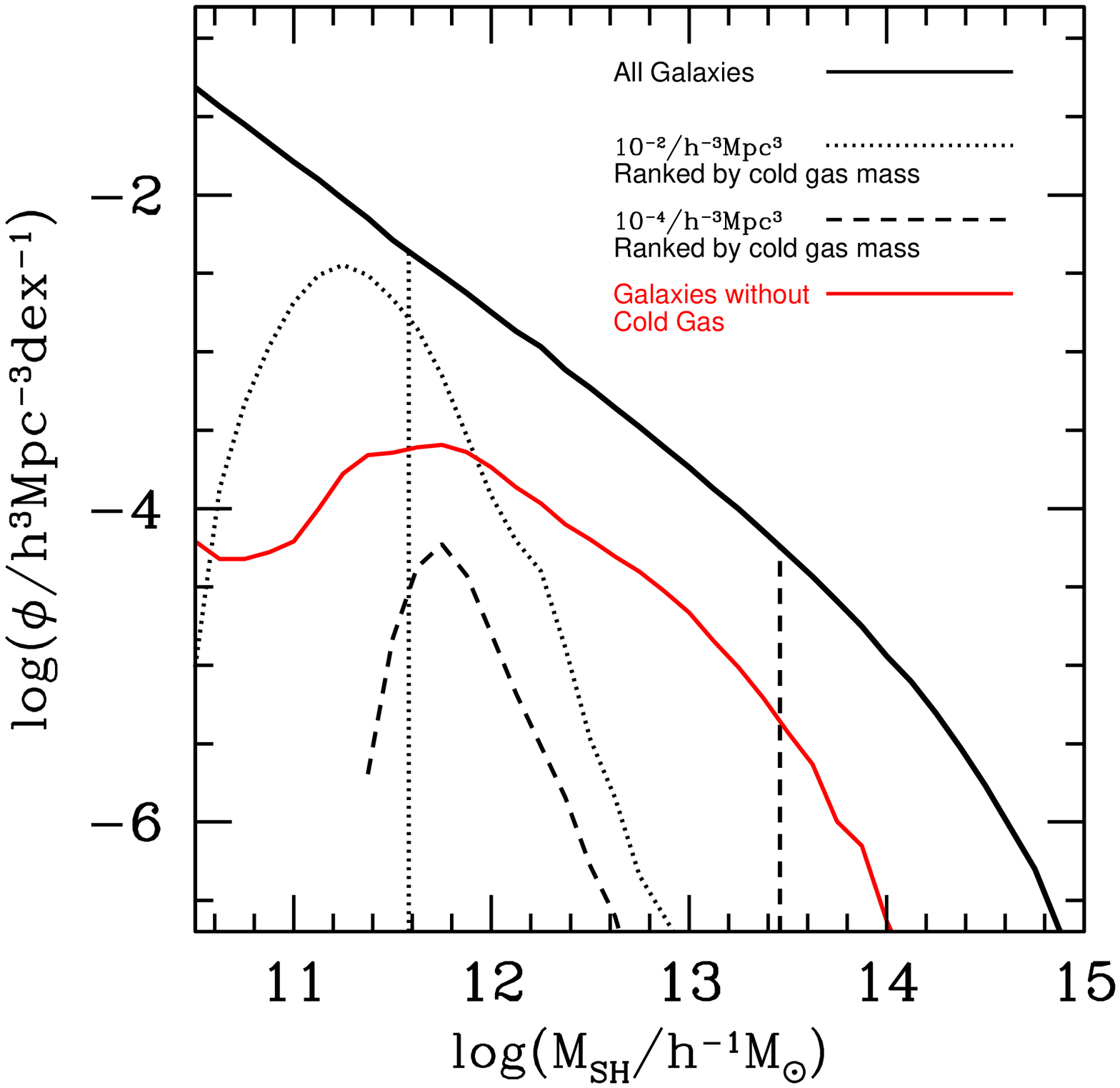}
\caption{
The subhalo mass function in the L12 model at $z=0$, constructed using 
all galaxies output by the model (solid black line in both panels). 
The vertical lines mark the masses above which the subhalos have abundances of 
$10^{-4} h^{3} {\rm Mpc}^{-3}$ (dashed line) and $10^{-2} h^{3} {\rm Mpc}^{-3}$ 
(dotted line). The other lines show the distribution of subhalo masses 
associated with the galaxies which pass a given selection criterion. 
In the top panel, the subhalo mass function is plotted for galaxies ranked 
in order of decreasing stellar mass, for an abundance of 
$10^{-4} h^{3} {\rm Mpc}^{-3}$ (dashed line) and $10^{-2} h^{3} {\rm Mpc}^{-3}$ 
(dotted line). In the bottom panel the same lines are used to show the 
subhalo mass function for the same galaxy number densities, but this time 
the galaxies have been ranked in terms of their cold gas mass. 
The solid red line shows the subhalo mass function for galaxies 
without any cold gas. 
}
\label{fig:Halo_Cut} 
\end{figure}

We now present the model predictions for how different galaxy properties 
depend on the mass of their host halo (\S~\ref{ssec:subhalo-property}), 
before looking more carefully into which halos contribute galaxies to 
different number density samples (\S~\ref{ssec:which-subhalo}). We then 
illustrate these dependencies further by attempt to reconstruct the 
SAM output by using the basic SHAM scheme (i.e. a subhalo abundance 
matching scheme without scatter) and a related approach (\S 4.3).
Finally, in \S4.4 we examine SHAM reconstruction 
at high redshift. 

\subsection{Subhalo mass - galaxy property distributions} 
\label{ssec:subhalo-property}

We start by considering the predicted dependence of galaxy luminosity 
on host dark matter subhalo mass in the L12 model in Fig.~\ref{fig:2D_Mr}. 
Galaxy luminosity in the optical was the original suggestion for a property 
that might display a monotonic dependence on halo mass 
\citep{ValeOstriker:2004}. The shading in Fig.~\ref{fig:2D_Mr} shows 
the abundance of galaxies as a function of their rest-frame $r$-band 
magnitude and host subhalo mass. As discussed in the previous section, 
we show predictions obtained from the MS-I and MS-II N-body runs, 
with the black dashed lines marking the transition from one set of 
results to the other, as labelled. The points and lines show the 
median $r$-band magnitude in bins of subhalo mass. The $r$-band magnitude 
shows a steep dependence on halo mass up to a mass of $\approx 10^{11.5} h^{-1} 
{\rm M_{\odot}}$. Beyond this mass the median $r$-band magnitude brightens 
less rapidly with increasing subhalo mass. This change in the slope of 
the median luminosity can be traced back to the onset of AGN heating of 
the hot gaseous halo, which stops gas cooling in halos more massive 
than $\approx 10^{11.5} h^{-1} {\rm M_{\odot}}$. Remarkably, there is 
essentially no difference in the median luminosity - halo mass relation 
when restricting attention to only central or satellite galaxies. 
The same trends are seen at $z=1$ and $z=4$. 

\begin{figure*}
\includegraphics[bb=27 416 355 709, width=1.0\textwidth]{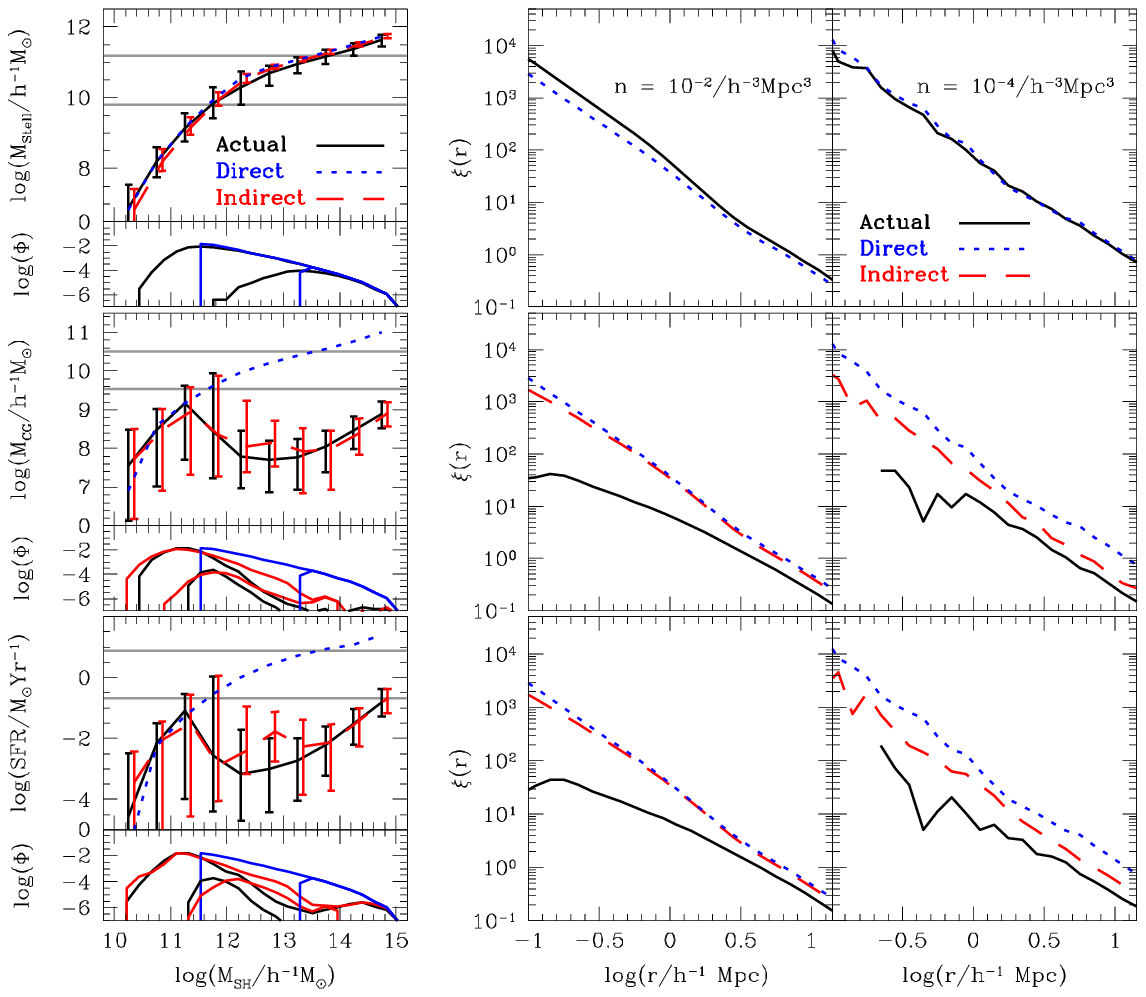}
\caption{ Tests of the accuracy of the reproduction of the actual galaxy 
sample predicted by the L12 model at $z=0$ using the direct and indirect 
SHAM reconstructions (see text). Each row shows the comparison for a 
different galaxy property (top - stellar mass; middle - cold gas mass; 
bottom - SFR). The main panels in the left column show the galaxy 
property - subhalo mass plane. The lines show the median galaxy property as 
a function of subhalo mass for the actual, direct and indirect samples as 
labelled. The lines showing the indirect samples have been shifted 
slightly for plotting clarity. The indirect curve is not shown in the 
top panel (stellar mass) since this is the same as the direct curve in 
this case.
The 20-80$^{\rm th}$ percentile range is shown for the actual and 
indirect samples. The horizontal lines mark the property values in the actual 
sample which define the high (lower line) and low (upper line) density 
samples: it is the region of the plane above these lines which is of interest 
for these samples. The lower sub-panels show the distributions of subhalo masses 
in these cases. The second and third columns show the correlation 
functions measured for the samples as labelled for high and low densities, 
respectively. 
}
\label{fig:Corr1} 
\end{figure*}

The {\it median} galaxy luminosity - halo mass relation satisfies the 
central assumption behind SHAM, showing a monotonic dependence on host 
halo mass. However, Fig.~\ref{fig:2D_Mr} shows that there is considerable 
scatter when individual galaxies are considered. The 20-80$^{\rm th}$ 
percentile range covers almost two magnitudes at the subhalo mass where 
the relation changes slope. The full range of galaxy magnitudes predicted 
in the model is much wider, covering around $8$ magnitudes or a factor of 
1500 in luminosity at the same mass. Similar results are found in other 
passbands. At longer wavelengths, galaxy luminosity is more closely related 
to stellar mass and the scatter in the luminosity - halo mass relation 
reduces slightly. At shorter wavelengths, the luminosity is driven more 
by the recent star formation history and also by the dust extinction, 
resulting in a more complicated dependence of luminosity on halo mass 
(see the discussion of SFR and luminosities at high-redshift in \S~4.4).  

Next we address the issue of the robustness of the SAM predictions 
by comparing the G11 and L12 models for different properties 
in Fig.~\ref{fig:2D}. Here we focus on physical galaxy properties; 
stellar mass, cold gas mass, black hole mass and star formation rate. 
The left and middle columns compare the predictions of G11 and L12 
respectively at $z=0$. If we first take the cases of stellar mass (top row) 
and black hole mass (third row down), the overall trends predicted by the 
two SAMs are similar, with more scatter predicted in the {\tt GALFORM} case.
{\tt GALFORM} also predicts a higher scatter than {\tt L-GALAXIES} 
when we consider galaxy luminosity. This is due to differences in the 
assumptions made to model galaxy formation physics, such as the choice 
of the time available for gas to cool from the hot halo. 
Observationally, the scatter in the halo mass - central galaxy luminosity 
relation has been studied using the dynamics of satellite galaxies 
\citep{More:2009,More:2011}. However, the question of whether or not the 
scatter predicted by either model is inconsistent with such observations 
remains open, as a careful comparison is required, repeating the 
analysis applied to the observations on a mock galaxy catalogue derived 
from the semi-analytical models, which is beyond the scope of the current 
paper. For both models, the stellar mass - halo mass relation 
changes slope at the 
halo mass at which AGN feedback starts to become important. Even though 
the models were calibrated to fit different datasets (primarily the stellar 
mass function in the case of G11 and the optical and near-infrared 
luminosity functions for L12), the change in slope occurs at approximately 
the same subhalo mass. 

The predicted distributions for cold gas mass and star formation rate 
are closely related in G11, where all of the cold gas mass above some 
critical value is made available for star formation. In L12, only 
molecular hydrogen takes part in star formation, so there is no longer 
a direct link between the total cold gas mass and the star formation rate. 
Qualitatively, the cold gas - halo mass and SFR - halo mass distributions 
are similar for a given model. The distributions show the same 
features between models but are different in detail. At low halo masses, 
there is a reasonable correlation between cold gas and SFR and halo mass. 
This breaks down above the halo mass for which AGN feedback is important. 
The severity of the break is different in G11 and L12. This is because 
AGN feedback shuts down gas cooling completely in sufficiently massive 
halos in the L12 model, whereas the suppression of cooling is more gradual 
in G11. The relations between cold gas mass or SFR and subhalo mass are 
also different for central and satellite galaxies. Satellite galaxies are 
predicted to have lower median cold gas masses than centrals, with the 
difference being greater in L12 than in G11. This can be readily understood 
in terms of the differences in the treatment of cooling in satellites in 
the models. In L12, there is complete stripping of the hot halo when a 
galaxy becomes a satellite. In G11 the stripping of the hot gas is partial 
depending on the ram pressure experienced by the satellite as it orbits 
within the more massive halo. 

Fig.~\ref{fig:2D} also shows the evolution of the galaxy property - halo 
mass distributions between $z=0$ and $z=1$ in the L12 model. There is little 
change in these distributions over this time interval. Although the 
abundance of massive dark matter halos changes appreciably between $z=4$ 
and $z=0$, the fraction of mass contained in halos with masses typical of 
those which host galaxies shows little change over this period \citep{Mo:2002}. 
\begin{figure}
\includegraphics[bb=20 150 570 700, width=0.47\textwidth]{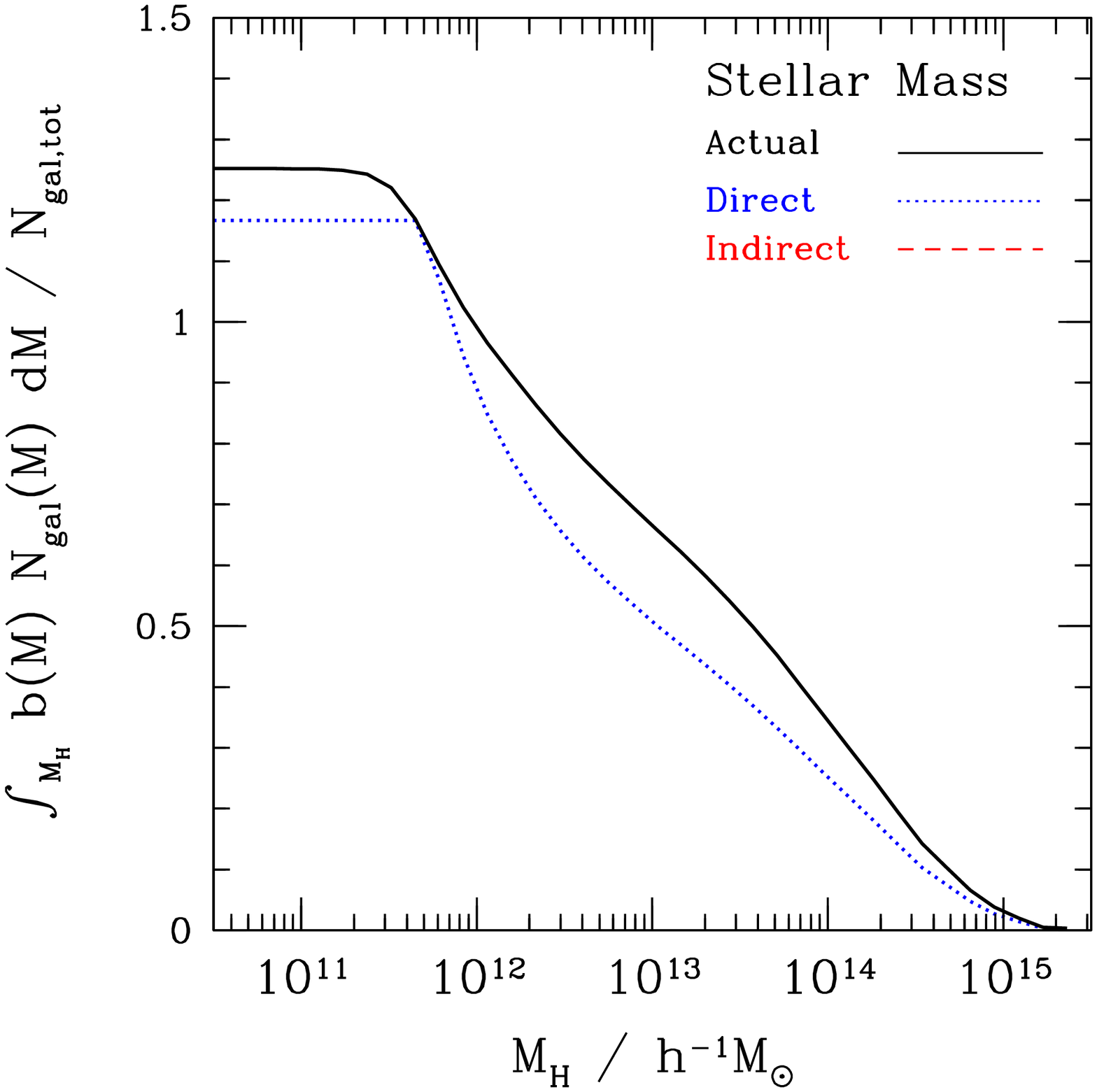}
\includegraphics[bb=20 150 570 700, width=0.47\textwidth]{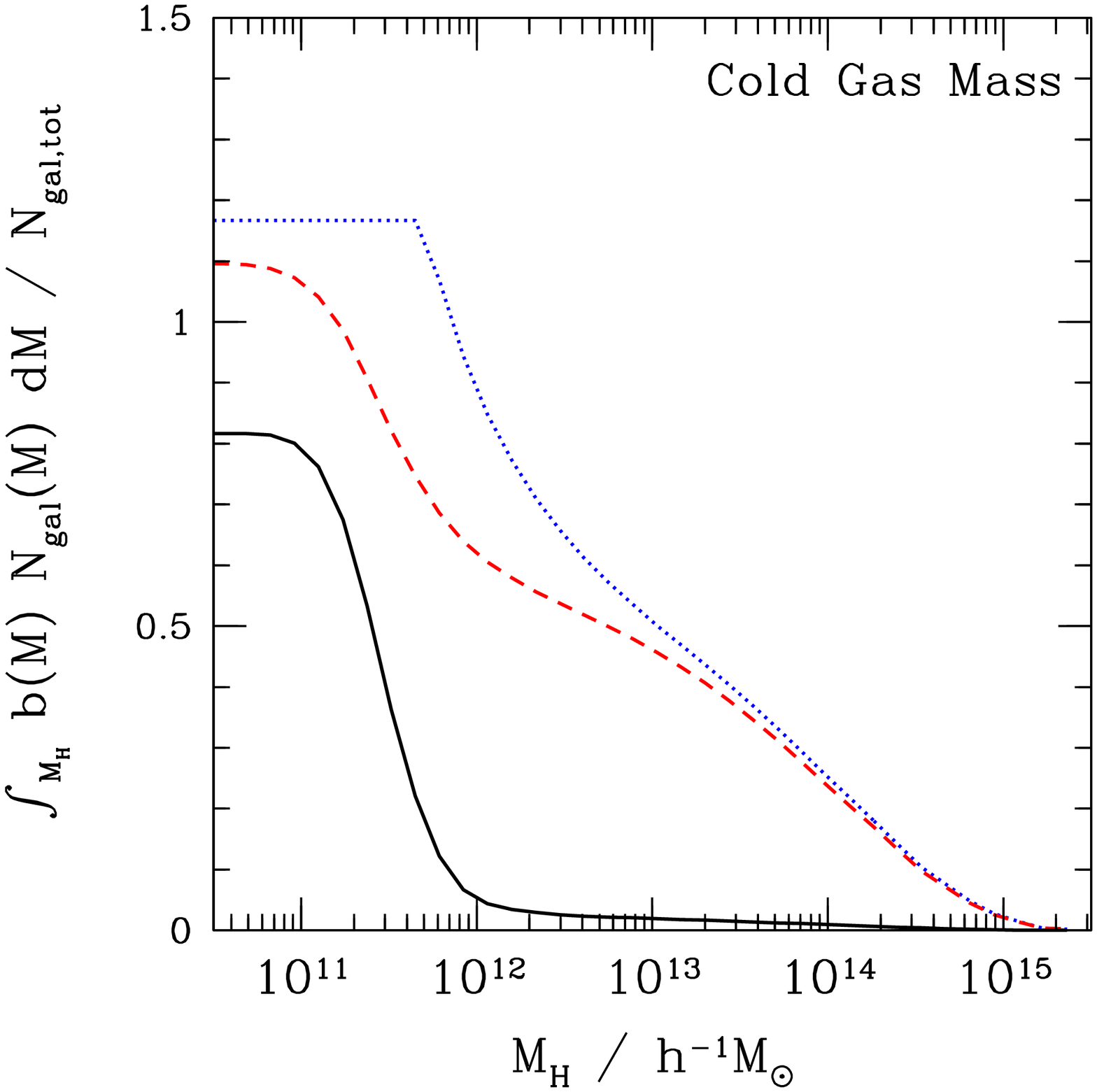}
\caption{
The contribution of different halos to the effective bias plotted 
as a function of halo mass, for galaxy samples with a number 
density of $10^{-2} h^{3} {\rm Mpc}^{-3}$ when ranked by their stellar mass 
(top panel) and by cold gas mass (botom panel)
The black curve shows the bias in the actual sample. 
The asymptotic bias value at low halo masses gives the effective bias 
of the sample as $b_{\rm eff} = 1.25$ for stellar mass and 
$b_{\rm eff} = 0.82$ for the stellar mass sample. 
For the stellar mass case the SHAM reconstruction (blue curve) 
gives an effective bias of $b_{\rm eff} = 1.18$ which is smaller 
than the actual bias. Note that there is no indirect curve in 
the top panel, since it has the same shape as the direct curve.
The bottom panels shows that the effective bias for the 
reconstruction of the cold gas selected sample is higher than 
the actual effective bias, in agreement with the correlation function 
results shown in Fig.~\ref{fig:Corr1}. The direct reconstruction 
effective bias curve flattens off at a halo mass of 
$4.5 \times 10^{11} h^{-1}M_{\odot}$ since there no halos with 
masses below this in the direct sample. 
}
\label{fig:bias} 
\end{figure}

\subsection{Which subhalos contain galaxies?}
\label{ssec:which-subhalo}
\begin{figure*}
\includegraphics[bb=27 416 355 709, width=1.0\textwidth]{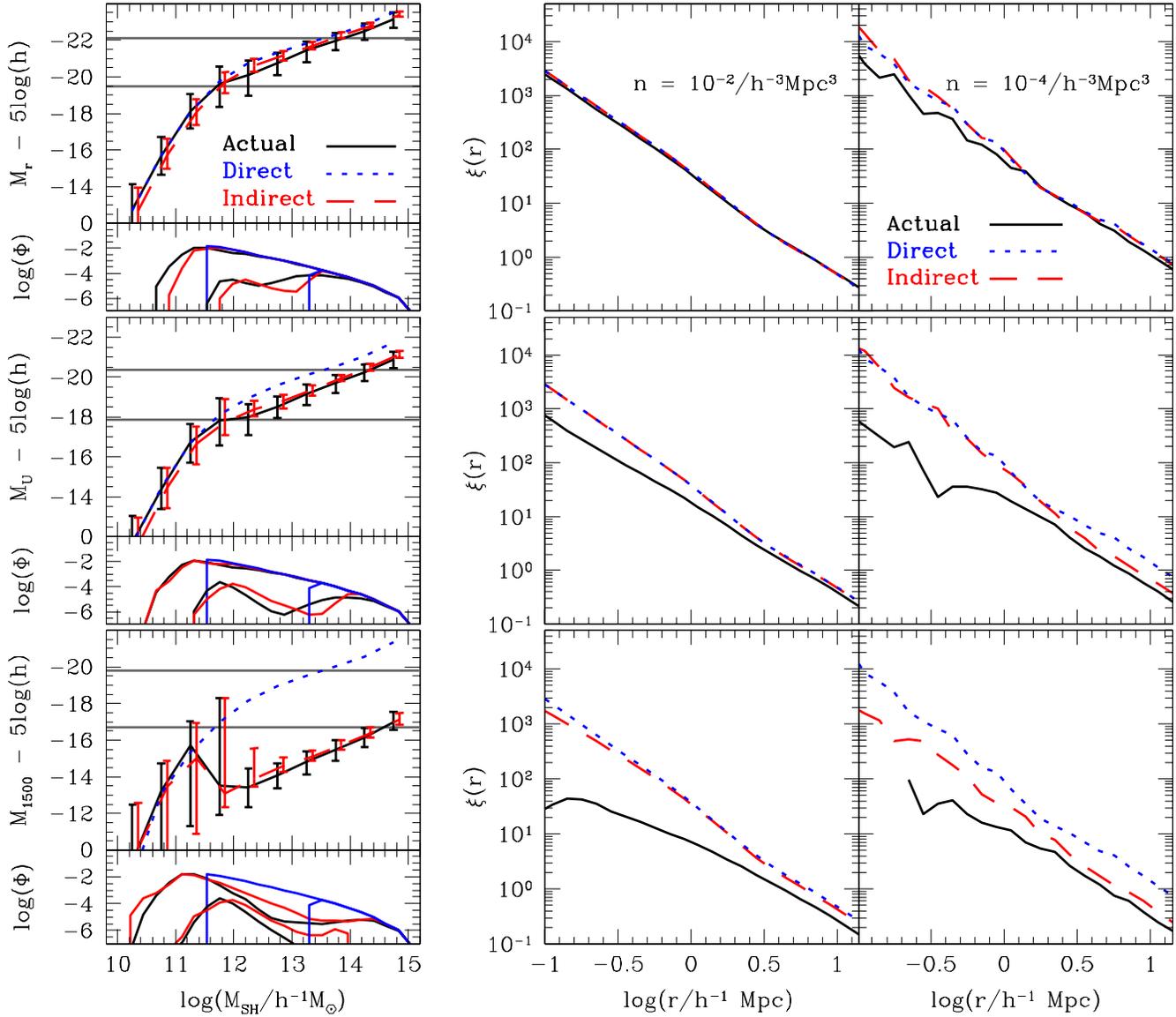}
\caption{
Same as Fig.~\ref{fig:Corr1}, but this time the galaxy samples are defined by the 
magnitude in different bands: $r$-band (top), $U$-band (middle) and 
$1500$\AA~(bottom).}
\label{fig:Corr2} 
\end{figure*}

In the previous subsection we showed how galaxy properties are 
predicted to depend on subhalo mass. All of the properties considered 
display an appreciable scatter for a given halo mass. For some properties, 
such as cold gas mass and SFR, the dependence on subhalo mass is complex, 
which means that these galaxy properties are not good indicators of host 
halo mass. In this subsection we demonstrate the features of the 
model predictions by applying the basic SHAM hypothesis to 
reconstruct the SAM catalogues. We show the impact of this simple
SHAM reconstruction by examining the range of halo masses populated 
with galaxies compared to that in the original catalogues, and the effect on 
the galaxy correlation function. 

To gain some insight into the results presented later on in this section, 
we first examine which parts of the overall subhalo mass function are 
represented when different galaxy selections are made. Fig.~\ref{fig:Halo_Cut} 
shows the subhalo mass function for subhalos associated with different galaxy 
samples for the L12 model at $z=0$. The solid black line shows the mass 
function when using the subhalos associated with all of the galaxies in 
the model output. This is our estimate of the ``true'' or complete 
subhalo mass function. We then build subsamples of galaxies by ranking them 
in terms of decreasing stellar mass (top panel) or cold gas mass 
(bottom panel) and plot the mass function of the associated subhalos. 
We do this for two galaxy number densities, $10^{-4} h^{3} {\rm Mpc}^{-3}$ 
(dashed lines) and $10^{-2} h^{3} {\rm Mpc}^{-3}$ (dotted lines). If a galaxy 
property satisfied the basic SHAM hypothesis exactly, then the mass 
function of 
the associated subhalos would include all of the available subhalos down to 
some mass, with a sharp transition to include zero subhalos of lower masses. 
This is indicated by for the two number densities by the vertical dashed 
and dotted lines in Fig.~\ref{fig:Halo_Cut}.

When galaxies are ranked in terms of their stellar mass, 
Fig.~\ref{fig:Halo_Cut} shows that all of the subhalos above some mass 
are selected (e.g. a halo mass of $10^{12.6} h^{-1} {\rm M_{\odot}}$ 
for a galaxy abundance of $10^{-2} h^{3} {\rm Mpc}^{-3}$). 
However, due to the steepness of the halo mass function, the samples are 
dominated by somewhat lower halo masses, around 
$10^{11.4} h^{-1} {\rm M_{\odot}}$ in this case. At this mass, roughly 
half of the available subhalos are predicted to contain a galaxy which 
satisfies the cut in stellar mass which defines the sample. There is a 
tail of lower mass halos, extending roughly an order of magnitude in mass 
below the peak which also contribute. In these halos there is a declining 
chance (dropping to 1 in 10000 for the range of masses shown by the 
dotted line) that the halo contains a sufficiently massive galaxy. 

The situation is more complex when galaxies are ranked by their cold gas 
mass. Fig.~\ref{fig:Halo_Cut} shows that for both number density cuts, only 
a very small fraction of massive halos are represented. The peaks of the 
mass functions shown by the dotted and solid lines lie far below the overall  
subhalo mass function. This means that even for the most common subhalo mass 
present in the sample, only 1 in 3 halos (for the sample with space density 
$10^{-2} h^{3} {\rm Mpc}^{-3}$) or 1 in 100 halos (for the 
$10^{-4} h^{3} {\rm Mpc}^{-3}$ sample) make it into these catalogues. 
In the case of cold gas, it is much more likely that a massive subhalo (i.e. 
with mass $>10^{12} h^{-1} {\rm M_{\odot}}$) will contain galaxies with no cold 
gas (red line) than with enough cold gas to be selected. The presence of a 
sizable population of subhalos without cold gas is supported by a recent 
interpretation of the clustering strength of H\,I selected 
samples \citep{Papastergis:2013}. Hence cold gas is not a suitable property to 
use in a direct basic SHAM analysis.

\subsection{SHAM reconstruction of the SAM model predictions} 

We now apply the basic SHAM method (i.e. assuming no scatter in a galaxy property for a given subhalo mass) to 
reconstruct the L12 galaxy catalogue. We compare three types of galaxy catalogues
as listed below: 
\begin{itemize}
\item{\bf Actual:} This is the catalogue {\it predicted} by the L12 SAM. 
Galaxies are ranked in terms of the galaxy property under consideration, 
in descending order of the property value. Two samples are used, 
corresponding to high ($10^{-2} h^{3} {\rm Mpc}^{-3}$) and 
low ($10^{-4} h^{3} {\rm Mpc}^{-3}$) space densities, corresponding to
$1.25 \times 10^{6}$ and $1.25 \times 10^{4}$ galaxies respectively for
the models run with MS-I. 
\item{\bf Direct:} This is a reconstruction of the actual sample using 
the basic SHAM approach. The entire actual catalogue is effectively 
used to generate two ranked order lists: one ordered in terms of declining 
subhalo mass and the other in terms of the galaxy property under 
consideration. Galaxies are then assigned a subhalo mass determined by their 
position in the rank-ordered list i.e. the galaxy with the largest property 
value is assigned to the most massive subhalo and so-on down the list until 
the desired space density is attained. 
\item{\bf Indirect:} This is a two step process in which SHAM is first 
applied to obtain the galaxy stellar mass. In the second step the target galaxy 
property is assigned by drawing from the distribution of the property 
as a function of stellar mass as predicted by the SAM (see \citealt{RP:2011}). 
In practice, when the galaxies are sorted in terms of their stellar mass, 
the associated values of the other galaxy properties predicted by the model 
are remembered. We then assign the value of a particular property 
that is associated with the galaxy, given its position in the list 
that is rank-ordered in terms of stellar mass.  
This approach can also be used to include scatter in the predicted 
galaxy property - subhalo mass distribution (though not in the case 
of the stellar mass, unless a different property is used in the 
first SHAM step to generate a rank-ordered list). By construction, for
galaxies selected by stellar mass, the indirect and the direct samples
will be identical.

The main motivation for introducing the indirect approach is to
improve the reproduction of the distribution of galaxies in the
galaxy property -  subhalo mass plane, particularly for galaxy
properties which have a complex dependence on subhalo mass,
such as the cold gas mass.

The clustering signal in different samples is presented as
an illustration of how the reconstruction method changes
the relation between galaxies and their host dark matter halos
(ie the main subhalo in the case of satellite galaxies). This is
a challenging test of the reconstruction, as applying SHAM blurs
any relation that is present in the semi-analytical model output
between galaxy properties and local density, as a subhalo loses 
memory of whether it was originally a subhalo within a more massive
halo or an isolated halo.
\end{itemize}  

Here we have an advantage over studies which apply SHAM to reproduce 
galaxy clustering measured from observations in that we know the true 
or actual (using the terminology introduced above) subhalo mass attached 
to each galaxy, as predicted by the SAM. We judge how well the reproduction 
works by comparing the mass function of subhalos in the reconstructed sample 
to that in the actual sample and also by comparing the galaxy correlation 
function. If the reconstruction puts galaxies into the correct subhalos 
(i.e. those originally predicted by the SAM) then the galaxy correlation 
function will match that of the actual sample. 

The tests of the quality of the reproduction of the L12 model are shown in 
Fig.~\ref{fig:Corr1} where each row shows the results for a different 
physical property (top row - stellar mass, second row - cold gas mass, 
third row - star formation rate). The main panels in the left-hand column 
show the galaxy property - halo mass distribution, as quantified through 
the median property values and the 20 - 80$^{\rm th}$ percentile distribution. 
The medians are shown for each catalogue: actual, direct and indirect. 
The horizontal lines in these panels show the minimum property values 
that define the two actual samples: the upper line is for the low space 
density sample and the lower line is for the high space density case. 
This shows which part of the galaxy property - subhalo mass plane 
contributes to these samples. The lower sub-panel in the left column of 
Fig.~\ref{fig:Corr1} shows the distribution of subhalo masses attached 
to the galaxies in each sample. Finally, the other columns show the 
comparison of the two-point galaxy correlation function for the high 
density (middle) and low density sample (right). 

Starting with stellar mass (top panel), Fig.~\ref{fig:Corr1} shows that 
the direct 
SHAM approach gives a reasonable reproduction of the median stellar mass 
in the actual sample, returning median stellar masses that agree well with 
those in the actual sample for halo masses below 
$10^{12} h^{-1} {\rm M_{\odot}}$ and that are $\approx 0.2$ dex too high at 
higher subhalo masses. Note that for the stellar mass the indirect curve is not
shown since it is identical to the direct curve; the median relation for the 
indirect method is plotted using bins that have been shifted slightly for 
clarity. The width of the distribution 
of stellar masses is smaller in the reconstructed samples than in 
the ``actual'' catalogue. The actual sample contains galaxies in lower mass 
subhalos than the simple SHAM reconstructions. 

\begin{figure}
\includegraphics[width=0.45\textwidth]{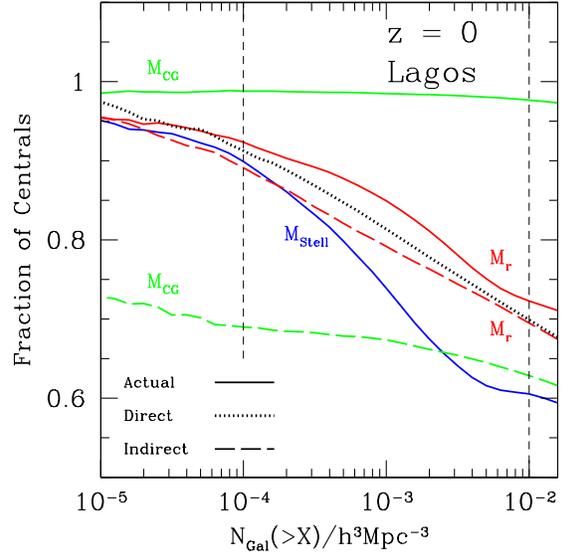}
\caption{
The fraction of central galaxies for different galaxy selections  
plotted as a function of the number density of galaxies in the sample, 
which corresponds to reducing the value of the property which 
is used to define the sample. 
The dotted black line shows the fraction of ``central" subhalos in the 
direct sample. Different line colours and styles refer to different 
selections as indicated. 
}
\label{fig:Sat} 
\end{figure}

Out of all the properties we have studied, stellar mass is the only 
one for which the direct SHAM reconstruction leads to an {\it underprediction} 
of the correlation function (for the $10^{-2} h^3Mpc^{-3}$ density cut). 
In this case, the direct approach puts galaxies into lower mass 
subhalos than in the actual sample, as shown in the top panel of 
Fig.~\ref{fig:bias}. This behaviour is critically dependent on the  
fraction of the subhalos of a given mass that are occupied, as shown 
in Fig.~\ref{fig:Halo_Cut}. 

We now consider the impact of the reconstruction on the predicted clustering. 
The relevant part of the stellar mass - subhalo plane to focus on now is that 
above the horizontal lines in the top-left panel. In this case, all three 
catalogues show very similar distributions of subhalo masses (as shown by 
the lower left panel in this row). The clustering predictions are extremely 
close to one another for the low density sample. For the high density sample, 
the reconstructions predict a slightly lower clustering amplitude, with the 
discrepancy reaching $\approx 60\%$ on small scales. 

The reconstructions work less well in the case of samples defined by their 
cold gas mass, as shown by the second row of Fig.~\ref{fig:Corr1}. Applying 
the direct SHAM approach results in a monotonic relation between cold gas 
mass and subhalo mass. The predicted distribution in the actual sample 
is very different. There are three values of the subhalo mass compatible 
with a median cold gas mass of $\approx 10^{8} h^{-1} {\rm M_{\odot}}$ 
in the case of the actual sample. The direct approach puts galaxies into 
more massive subhalos than the model predicts. The indirect, two-step 
approach does a much better job of putting galaxies in subhalos of the correct 
mass and matching the width of the cold gas mass distribution. However, 
the clustering signal predicted by the reconstructions is much higher than 
the actual prediction, particularly for the high density sample. Remember, 
for the two samples under consideration we are only interested in the region 
of the cold gas - subhalo mass plane which lies above the horizontal lines. 
For the actual and indirect samples, the median cold gas mass is always 
below these lines, so we are focusing on the extremes of the distribution. 
Similar behaviour is found for the case of the SFR, as shown by the bottom 
row in Fig.~\ref{fig:Corr1}. 

Fig.~\ref{fig:Corr1} shows that the clustering in the reconstructions 
for samples defined by cold gas mass is higher than the prediction in 
the SAM. The contribution to the effective bias as function of halo 
mass is shown in the bottom panel of Fig.~\ref{fig:bias}. The curve for the ``actual'' sample 
is always below those for the reconstructions, which means that the 
reconstructions preferentially populate higher mass subhalos with galaxies 
than is the case in the actual sample. The difference in the effective 
bias between the samples matches the difference seen in the two-halo 
term in the correlation function in Fig.~\ref{fig:Corr1}. 

Fig.~\ref{fig:Corr2} shows the results of the reconstruction of 
samples defined by galaxy luminosity in different bands. The top 
row shows the $r$-band (effective wavelength $\lambda_{\rm eff}=6166$~\AA), 
the middle the $U$-band ($\lambda_{\rm eff} = 3509$~\AA) and the bottom 
row is for a rest frame wavelength of $1500$~\AA.
For the high density sample, the correlation function obtained 
from the reconstructions (direct and indirect) agrees well 
with that for the actual sample. For this density cut the $r$-band 
is the one the one that shows the best agreement with the direct 
reconstruction.
For the low density sample, the clustering in the reconstructions is 
somewhat higher than in the actual sample, particularly on small scales. 
The $U$-band is more sensitive to the SFR and also to the dust extinction 
in the galaxy. The direct reconstruction does not work well in this case 
for subhalos more massive than $10^{11.8} h^{-1} {\rm M_{\odot}}$, predicting 
a median galaxy magnitude that is around one magnitude brighter than in the 
actual catalogue for massive halos. The indirect approach fares better. 
Nevertheless, both reconstructions overpredict the amplitude of the correlation 
function. Fig.~\ref{fig:Corr2} shows that the largest discrepancy between 
the reconstructions and the actual sample is found in the far-ultraviolet 
at $1500$~\AA. The median magnitude has a nonmonotonic dependence on halo 
mass in the actual sample. This is reproduced reasonably well in the indirect 
reconstruction. However, by construction, this behaviour cannot be obtained 
from the direct approach. Neither of the reconstructions gives an accurate 
reproduction of the clustering in the actual sample. Although the indirect 
approach can reproduce the median magnitude - subhalo mass relation 
predicted in the actual sample, the number densities of galaxies under 
consideration means that it is the extreme of this distribution that is  
being probed in the clustering comparison. The reconstructions clearly do 
not reproduce the tails of the distributions.

Finally, we consider how the SHAM reconstruction affects 
the division between central and satellite galaxies. The number and 
spatial distribution 
of satellite galaxies in a halo shapes the form of the two-point correlation 
function on small scales and is referred to as the one-halo term. The 
largest differences seen in the correlation functions plotted in 
Fig.~\ref{fig:Corr1} and ~\ref{fig:Corr2} occur on the scales sensitive 
to the one-halo term. 

The SAM predicts which galaxies are centrals and which are satellites. 
In Fig.~\ref{fig:Sat} we show the fraction of central galaxies 
predicted in the L12 model at $z=0$, as a function of the number density 
of the sample, when selecting using different galaxy properties. For the 
actual sample (solid line), the 
fraction of central galaxies shows similar behaviour when selecting 
on stellar mass or $r$-band magnitude. At low galaxy number densities, 
the samples are dominated by centrals, with the low-density sample 
containing around 90\% centrals. The fraction of centrals drops with 
increasing galaxy number density, reaching 60\% for stellar mass selection 
and 72\% for $r$-band selection in the high density sample. However, 
when selecting on cold gas mass, the fraction of centrals is remarkably 
insensitive to the abundance of galaxies.   

The designation in the SAM of a galaxy as a ``central'' or ``satellite'' 
can also be used to label the subhalo hosting the galaxy. In the direct 
SHAM reconstruction, we can track the fraction of the ``central'' subhalos 
after the halos are rank-ordered in mass. This is shown by the black dotted curve 
in Fig.~\ref{fig:Sat}. This curve has a similar shape to that 
predicted by the SAM for stellar mass and $r$-band selection. Hence we would 
expect the direct SHAM reconstruction of galaxy samples defined by these 
properties to produce similar numbers of central and satellite galaxies as 
predicted in the actual sample. This is not the case for cold gas selection, 
with the direct SHAM reconstruction predicting many more satellites than the 
model contains (comparing the black and green lines in Fig.~\ref{fig:Sat}). 

The dashed line in Fig.~\ref{fig:Sat} shows the fraction of centrals in 
the indirect SHAM reconstructions. The fraction of centrals in the $r$-band 
reconstruction is slightly lower than in the actual L12 predictions, but 
shows a similar trend with galaxy number density. The sample reconstructed 
using cold gas mass shows a much lower fraction of central galaxies, indicating 
that the indirect SHAM puts more galaxies into subhalos which were originally 
satellite subhalos, instead of putting them into central subhalos. 
This boosts the amplitude of the one-halo term in the correlation function. This 
is consistent with the results shown for the effective bias of these samples in 
Fig.~\ref{fig:bias}.

\subsection{Applying SHAM at high-redshift}
\label{ssec:sham-highz}

\begin{figure*}
\includegraphics[bb= 26 598 373 710, width=1.05\textwidth]{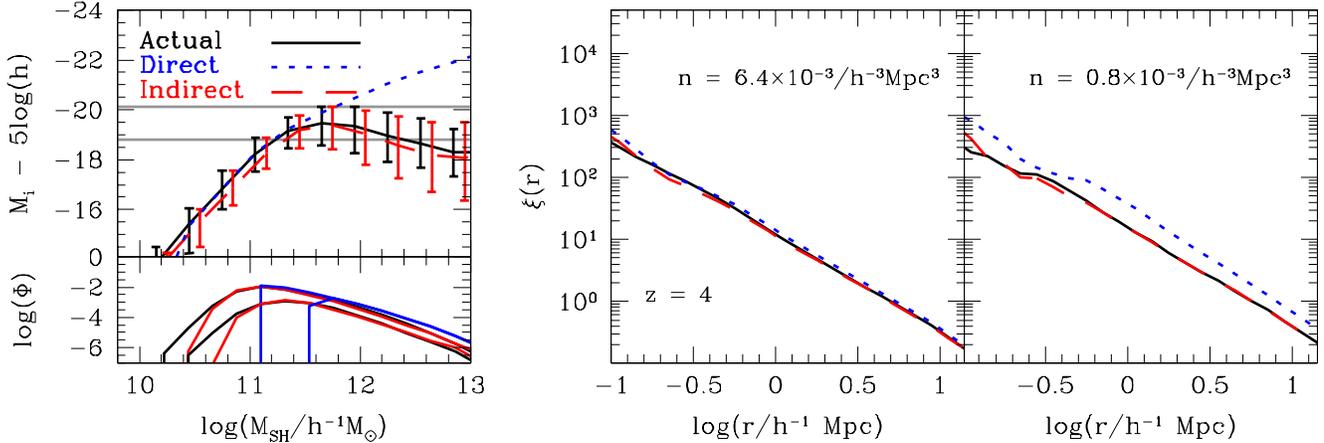}
\caption{
An application of SHAM at $z=4$, in a similar format 
to Figs.~\ref{fig:Corr1} 
and~\ref{fig:Corr2}. Here galaxies are ranked by their observer-frame $i$-band 
magnitude. Density cuts of $0.8 \times 10^{-3} h^{-3} {\rm Mpc}^{3}$ 
and $6.4 \times 10^{-3} h^{-3} {\rm Mpc}^{3}$ are used to match the sample 
selection adopted by Conroy et~al. (2006). The main panel in the left column 
shows the distribution in the $i-$band magnitude - subhalo mass plane, with 
the lower panel showing the abundance of haloes in each sample. The middle 
and right panels show the correlation function in the two galaxy samples, 
as labelled.}
\label{fig:Co6} 
\end{figure*}

We now apply the basic SHAM scheme to reconstruct 
the L12 model predictions at $z=4$. 
The objective is to test the application of the basic SHAM 
technique to model 
the clustering of Lyman-break galaxies used by \cite{Conroy:2006}. 
The observational sample considered by Conroy et~al. was selected in the 
observer-frame $i$ band, which, at this redshift probes an effective rest-frame 
wavelength of $\approx 1600$\AA. The third row of Fig.~\ref{fig:Corr2} shows 
a similar test at $z=0$ and indicates that galaxy luminosity in the far 
ultra-violet is not a suitable property to use in a basic SHAM 
scheme, unless a fortuitous 
choice of galaxy number density is made. The comparison of the actual 
sample and the SHAM reconstructions is shown in Fig.~\ref{fig:Co6}, which 
is in the same format as Figs.~\ref{fig:Corr1} and~\ref{fig:Corr2}. 
The left panel of Fig.~\ref{fig:Co6} shows that the L12 model predicts a 
non-monotonic dependence of $i$-band magnitude on subhalo mass. 
The direct SHAM reconstruction overpredicts the brightness of galaxies 
hosted by massive halos. The upper of the two horizontal lines in 
Fig.~\ref{fig:Co6} shows that this will be a problem for the low-density 
sample. The indirect approach reproduces the median $i$-band magnitude 
as a function of halo mass much better, albeit with a slightly larger 
scatter for massive subhalos. The lower panel shows that the direct 
SHAM puts galaxies into more massive subhalos than predicted by the 
actual sample. The right panel shows that this results in an 
overprediction of the clustering amplitude using the direct SHAM 
approach for the low density sample. The indirect approach, 
on the other hand, gives a good reproduction of the actual clustering. 
The SHAM reconstructions both reproduce the clustering in the actual 
catalogue for the higher number density sample. The left panel of 
Fig.~\ref{fig:Co6} shows why this is the case. The lower of the two 
horizontal lines shows the $i$-band magnitude which is 
the selection limit for the high number density sample. 
This line intersects the actual, direct and indirect 
curves at the same place. Due to the steepness of the galaxy 
luminosity function and the subhalo mass function, it is this 
agreement which matters for the accuracy of the reproduction 
of the sample, as galaxies with luminosities close to this limit 
dominate. The disagreement between the actual and direct samples 
for higher subhalo masses does not matter in this case, as this 
only affects a small fraction of the overall sample. 

In summary, the direct basic SHAM approach will not work for low 
density galaxy samples when the relation between galaxy property 
and subhalo mass is not monotonic. If a sufficiently high number 
density sample is considered, then SHAM will work provided that 
galaxies in low mass subhalos dominate the sample (by number). 
A similar conclusion regarding the inappropriateness of applying 
SHAM to ultra-violet selected samples was reached in a study of 
close pairs of galaxies by \cite{Berrier:2012}.

\section{Summary and Conclusions}

We have explored the connection between the mass of 
dark matter subhaloes and the properties of the 
galaxies they contain, using physically motivated 
models of galaxy formation. If a simple, deterministic  
relation holds, this motivates the development of 
empirical models of the galaxy population, such as 
subhalo abundance matching (SHAM). 

The key assumption behind the original SHAM scheme (i.e. a scheme without scatter)  
is that there is a unique connection between a galaxy property 
and the mass of the galaxy's host dark matter halo. 
We have explored this assumption studing the galaxy - dark matter halo connection
in two independent, physically motivated models of galaxy 
formation. By using semi-analytical models implemented in the 
Millennium I and II N-body simulations \citep{Guo:2011,Lagos:2012}, we 
have been able to extend previous tests of SHAM well into the range of halo 
masses in which gas cooling is reduced by heating from active galactic 
nuclei \citep{Simha:2012}. This is a critical point as many of the 
most significant discrepancies from the basic SHAM assumption are found 
in massive halos. Another advantage of our study is the use of 
galaxy merger histories to track the mass of halos at the point of 
infall into a more massive halo. In this way we are able to include 
subhalos which are no longer identifiable in a single output of an 
N-body simulation. 

We have considered a range of intrinsic galaxy properties (stellar mass, 
cold gas mass, star formation rate, black hole mass) and direct 
observables (the luminosity in different bands, from the far ultra-violet 
to the optical). The model predictions show that {\it none} of these 
properties satisfy the basic SHAM assumption. Whilst some properties 
(stellar mass, black hole mass, $r$-band magnitude) display median values which 
vary monotonically with halo mass, a range of values is found for each 
halo mass. The models admittedly predict somewhat different ranges of 
property values, so the precise width of the distribution of values is 
a less robust model prediction. Some of this difference can be traced 
to choices made in the semi-analytical models (e.g. the definition of 
the time available for gas cooling). For other properties (cold gas mass, 
star formation rate, luminosity in the ultra-violet) the variation of 
the median  with halo mass is complex. For some property values in these 
cases, galaxies could appear in very different mass halos. 

The availability of the predictions of the galaxy formation models means 
that we can test how accurately SHAM can reconstruct the original catalogue. 
This exercise allows us to gain an impression of how the model predictions differ from the assumptions made in the simplest incarnations of SHAM.
If the real Universe looks like the galaxy formation models, then this 
process will inform us about possible systematic errors when using  
simple SHAM schemes to model observed galaxy clustering. 
We judge the quality of the reproduction in terms of the median and 
percentile range of galaxy property in bins of 
subhalo mass and in terms of the two-point galaxy correlation function. 
The direct SHAM reconstructions tend to put galaxies with too high a value 
of the property under consideration into massive subhalos. This in turn 
results in the clustering being too high in low-density galaxy samples, 
compared with the prediction in the model. The direct reconstruction fares 
better at lower subhalo masses, which are not affected by AGN feedback. 
Hence for high number density galaxy samples, which are dominated by 
galaxies in lower mass subhalos, SHAM tends to give a better reproduction 
of the predicted clustering. 

Extensions to the original SHAM proposal have been introduced to account 
for the scatter in the value of a galaxy property for a given subhalo 
mass and also to model properties which themselves are not thought to 
have a monotonic dependence on subhalo mass galaxy properties 
\citep{Tasitsiomi:2004,Behroozi:2010,Moster:2010,RP:2011,Hearin:2013b,
Gerke:2013,Masaki:2013,Hearin:2014,RP:2014}. Here we explore 
a simple variation on the basic SHAM scheme, which involves 
applying SHAM to one property and then assign galaxies a second property 
using a model which connects the two properties. The semi-analytical 
models predict the subhalo mass which hosts a galaxy, along with its 
intrinsic physical properties (e.g. stellar mass, cold gas mass, 
star formation rate). To build a sample for a property which does not 
have a simple dependence on host halo mass, we use a two-step approach. 
First, SHAM is applied to a galaxy property which does have a more 
straight forward relation to subhalo mass, as we found to be the case 
for stellar mass. Then to construct a sample which includes information 
about the desired galaxy property, for example the cold gas mass, we use the 
distribution of cold gas mass to stellar mass predicted by the 
semi-analytical model (see Appendix). We found this two step approach 
to be successful at reproducing the median and 20-80 percentile range 
of the target galaxy property as a function of subhalo mass. However, 
this approach does not always lead to the reproduction of the clustering 
signal in the model, particular for galaxy samples with a low number density. 
An extension to this approach could take into account the formation histories of the 
dark matter subhalos when assigning galaxy 
properties \citep{Gao:2005,Wang:2013}.

\section*{Acknowledgements}
This work was made possible by the efforts of Gerard Lemson and 
colleagues at the German Astronomical Virtual Observatory in setting up 
the Millennium Simulation database in Garching, 
and John Helly and Lydia Heck in setting up the mirror at Durham. 
We acknowledge helpful conversations with the participants 
in the GALFORM lunches in Durham. This work was partially 
supported by ``Centro de Astronom\'\i a y Tecnolog\'\i 
as Afines" BASAL PFB-06. NP was supported by Fondecyt 
Regular \#1150300. We acknowledge support from the European 
Commission's Framework Programme 7, through the Marie Curie 
International Research Staff Exchange Scheme LACEGAL (PIRSES-GA-2010-269264).
PN acknowledges the support of the Royal Society through the award 
of a University Research Fellowship and the European Research 
Council, through receipt of a Starting Grant (DEGAS-259586). 
PN \& CMB acknowledge the support of the Science and Technology Facilities 
Council (ST/L00075X/1). SC acknowledges support from CONICYT Doctoral Fellowship Programme. 
The calculations for this paper were performed on the 
ICC Cosmology Machine, which is part of the DiRAC-2 
Facility jointly funded by STFC, the Large Facilities 
Capital Fund of BIS, and Durham University and on the Geryon computer at the
Center for Astro-Engineering UC, part of the BASAL PFB-06, which received
additional funding from QUIMAL 130008 and Fondequip AIC-57 for upgrades.

\bibliography{Biblio}

\appendix

\section{Predictions for dependence of selected galaxy properties on stellar 
mass} 

\begin{figure}
\includegraphics[bb= 22 151 568 704, width=0.48\textwidth]{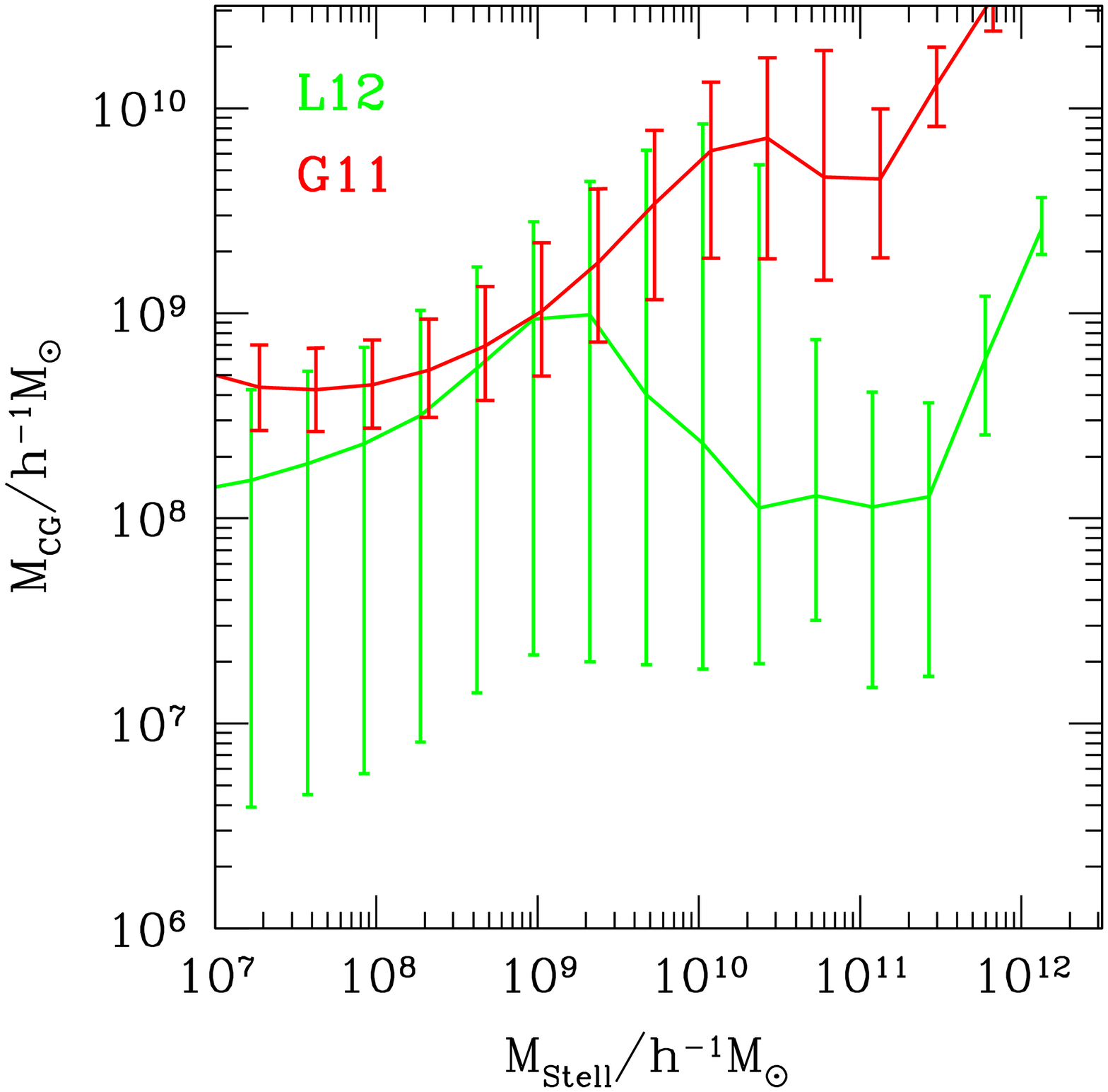}
\includegraphics[bb= 22 151 568 704, width=0.48\textwidth]{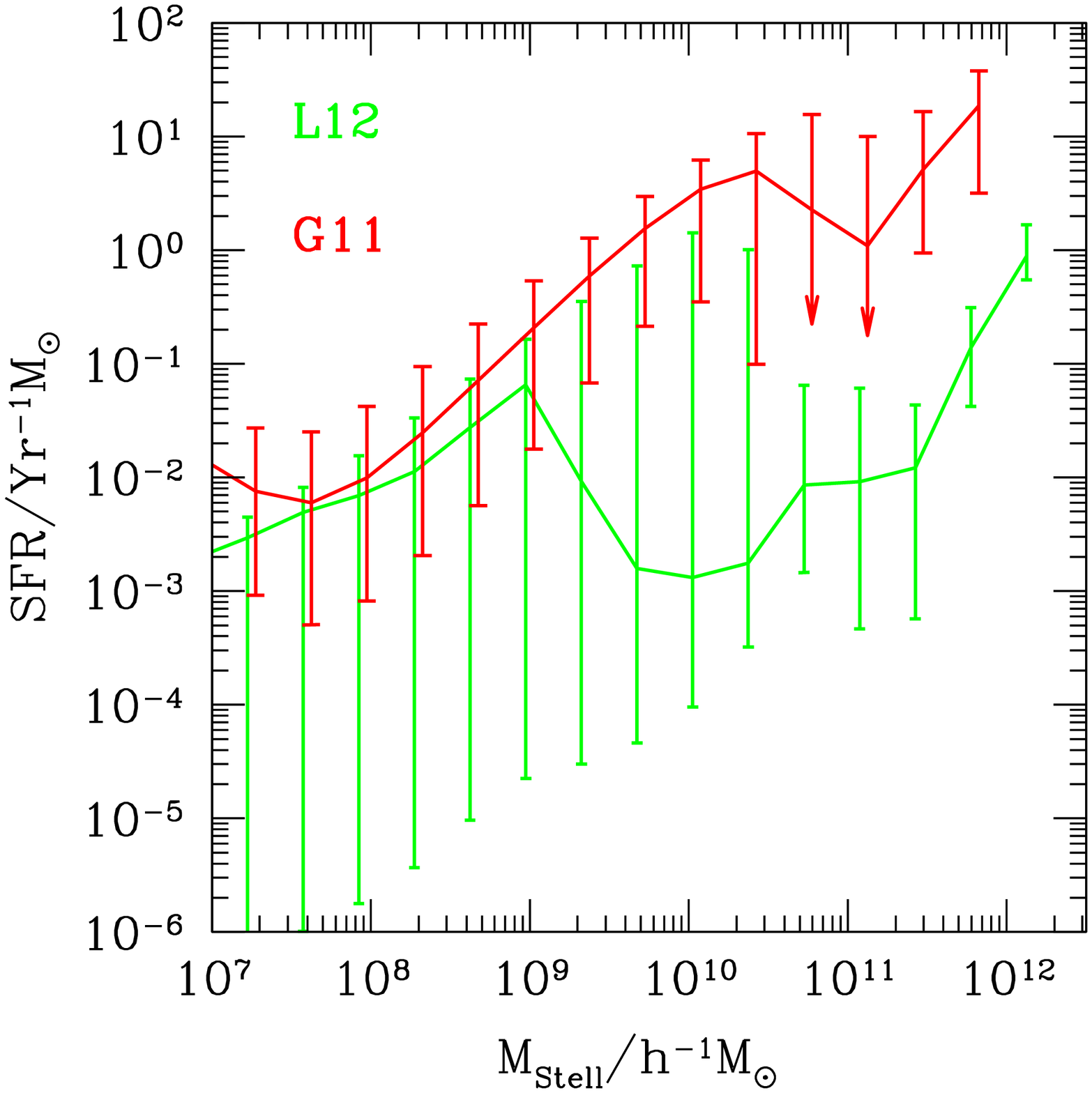}
\caption{The predicted dependence of cold gas mass (top) and star formation 
rate (bottom) on stellar mass in the G11 (red) and L12 (green) models. The 
lines show the median value and the bars show the 20-80\% percentile range. 
The downwards pointing arrow in the bottom plot means that the 20$^{\rm th}$percentile of the distribution has zero SFR.
}
\label{fig:app} 
\end{figure}

Motivated by the predictions of the galaxy formation model, we consider an indirect, two-step SHAM approach in which galaxies are assigned a 
property based on their stellar mass. This requires knowledge of how the 
desired or target galaxy property depends on stellar mass. 
Fig.~\ref{fig:app} shows the G11 and L12 model predictions for the 
dependence of cold gas mass (top) and star formation rate (bottom) 
on stellar mass. This information could be used in the indirect SHAM 
approach to build galaxy samples which cold gas information. 
Note that the models have not been calibrated to reproduce the same 
observations, hence the differences in these predictions. 
The L12 model predicts more scatter in cold gas mass and star 
formation rate for a given stellar mass than the G11 model.

\label{lastpage}

\end{document}